\documentclass[a4paper,12pt]{iopart}
\usepackage{iopams}
\usepackage{setstack}
\usepackage{graphicx}
\usepackage{bm}
\usepackage{epsfig}

\newcommand{\diag}{{\rm diag\,}}
\newcommand{\sign}{{\rm sign\,}}
\newcommand{\Str}{{\rm Str\,}}
\newcommand{\Sdet}{{\rm Sdet\,}}

\newcommand{\U}{{\rm U\,}}

\newcommand{\Herm}{{\rm Herm\,}}
\newcommand{\Ber}{{\rm Ber\,}}
\newcommand{\RE}{{\rm Re\,}}

\newcommand{\IM}{{\rm Im\,}}
\newcommand{\eins}{\leavevmode\hbox{\small1\kern-3.8pt\normalsize1}}

\begin{document}

\newtheorem{definition}{Definition}[section]
\newtheorem{assumption}[definition]{Assumption}
\newtheorem{theorem}[definition]{Theorem}
\newtheorem{lemma}[definition]{Lemma}
\newtheorem{corollary}[definition]{Corollary}

\title{On the Efetov-Wegner terms by diagonalizing a Hermitian supermatrix}
\author{Mario Kieburg}
\address{Universit\"at Duisburg-Essen, Fakult\"at f\"ur Physik, Lotharstrasse 1, 47048 Duisburg, Germany}
\eads{\mailto{mario.kieburg@uni-due.de}}

\begin{abstract}
The diagonalization of Hermitian supermatrices is studied. Such a change of coordinates is inevitable to find certain structures in random matrix theory. However it still poses serious problems since up to now the calculation of all Rothstein contributions known as Efetov-Wegner terms in physics was quite cumbersome. We derive the supermatrix Bessel function with all Efetov-Wegner terms for an arbitrary rotation invariant probability density function. As applications we consider representations of generating functions for Hermitian random matrices with and without an external field as integrals over eigenvalues of Hermitian supermatrices. All results are obtained with all Efetov-Wegner terms which were unknown before in such an explicit and compact representation.
\end{abstract}

\maketitle

\section{Introduction}\label{sec1}

Many eigenvalue correlations for random matrix ensembles as matrix Green functions, the $k$-point correlation functions \cite{Guh06,Zir06} as well as the free energy \cite{VerZir85} can be derived by generating functions. These functions are averages over ratios of characteristic polynomials for random matrices.

A common approach to calculate generating functions is the supersymmetry method \cite{LSSS95,EST04,Guh06,LSZ07}. Another approach is the orthogonal polynomial method \cite{Meh04}. In the supersymmetry method one maps integrals over ordinary matrices to integrals over supermatrices. Its advantage is the drastic reduction of the number of integration variables. Nevertheless there is a disadvantage. Up to now it is not completely clear how to get the full structures found with the orthogonal polynomial method for factorizing probability densities. For such probability densities the $k$-point correlation functions can be written as determinants and Pfaffians of certain kernels. This property was extended to the generating functions \cite{GGK04,BorStr06,KieGuh09a,KieGuh09b}. Unfortunately, the determinants and Pfaffians were not found for the full generating function after mapping the integrals over ordinary matrices to integrals over supermatrices. Only the determinantal expression of the $k$-point correlation function for rotation invariant Hermitian random matrix ensembles could be regained with help of the supersymmetry method \cite{Guh06,KGG08}. Gr\"onqvist, Guhr and Kohler studied this problem from another point of view in Ref.~\cite{GGK04}. They started from the determinantal and Pfaffian expressions of the $k$-point correlation functions for Gaussian orthogonal, unitary and symplectic random matrix ensembles and showed that the kernels of the generating functions with two characteristic polynomials as integrals in superspace yield the known result.

Changing coordinates in superspaces causes serious problems since the Berezinian \cite{Ber87} playing the role of the Jacobian in superspace incorporates differential operators. These differential operators have no analog in ordinary space and are known in the mathematical literature as Rothstein's vector fields \cite{Rot87}. For supermanifolds with boundaries such differential operators yield boundary terms. In physics these boundary terms are called Efetov-Wegner terms \cite{Efe83,Weg83,Efe97,KKG08}. They can be understood as corrections to the Berezinian without Rothstein's vector fields. There were several attempts to explicitly calculate these vector fields or the corresponding Efetov-Wegner terms for diagonalizations of Hermitian supermatrices but this was only successful for low dimensional supermatrices, e.g. $(1+1)\times(1+1)$ Hermitian supermatrices \cite{Guh93a,Guh93b,Guh93c}. For higher dimensions the calculation of all Efetov-Wegner terms becomes cumbersome. Also for other sets of supermatrices such changes in the coordinates were studied \cite{Zir91,Bun93,Zir96}. There are quite general formulas \cite{Rot87,Bun93,Pal10} for calculating the Efetov-Wegner terms but for concrete examples these formulas become cumbersome and one never came beyond the low dimensional cases. As an example for such an approach we tried to derive all Efetov-Wegner terms by considering these terms as boundary terms resulting from partial integrations of differential operators which are equivalent to the integration over the Grassmann variables \cite{KKG08}. Though a quite compact form for a differential operator for the Hermitian supermatrices was found we were unable to calculate all Efetov-Wegner terms. The order of such a differential operator is the number of pairs of Grassmann variables.

We derive the supermatrix Bessel function \cite{GuhKoh02b} for Hermitian supermatrices with all Efetov-Wegner terms. Thereby we apply a method called ``supersymmetry without supersymmetry'' \cite{KieGuh09a} on Hermitian matrix ensembles whose characteristic function factorizes in the eigen-representation of the random matrix. This approach uses determinantal structures of Berezinians without mapping into superspace. We combine this method with the supersymmetry method described in Refs.~\cite{Guh06,KGG08}. The supermatrix Bessel function is then obtained by simply identifying the left hand side with the right hand side of the resulting matrix integrals. As a simple application we calculate the generating function of arbitrary, finite dimensional, rotation invariant Hermitian matrix ensembles with and without an external field. Hence we generalize known results \cite{Guh06,Guh06b,KGG08,KieGuh09a} because they now contain all Efetov-Wegner terms which guarantee the correct normalizations.

We organize the article as follows. In Sec.~\ref{sec2}, we will give an outline of our approach and introduce some basic quantities. Using the method ``supersymmetry without supersymmetry'' we derive a determinantal structure for generating functions of rotation invariant Hermitian random matrices without an external field and with a factorizing characteristic function, in Sec.~\ref{sec3}. In Sec.~\ref{sec4}, we briefly present the results of the supersymmetry method. We also calculate the supermatrix representation of the generating function with one determinant in the numerator as in the denominator. For this generating function the corresponding supermatrix Bessel function with all Efetov-Wegner terms is known \cite{Guh91} since it only depends on $(1+1)\times(1+1)$ Hermitian supermatrices. In Sec.~\ref{sec5}, we plug the result of Sec.~\ref{sec4} into the result of Sec.~\ref{sec3} and obtain the supermatrix Bessel function with all Efetov-Wegner terms for arbitrary dimensional Hermitian supermatrices. Moreover we discuss this result with respect to double Fourier transformations and, thus, Dirac-distributions in superspace. In Sec.~\ref{sec6}, we, first, apply our result on arbitrary, rotation invariant Hermitian random matrices without an external field and, then, in the presence of an external field. Details of the calculations are given in the appendices.

\section{Outline}\label{sec2}

We want to generalize the well known result for Gaussian superfunctions \cite{Guh91,KKG08}
\begin{eqnarray}
 \fl\int \exp\left(-\frac{1}{2}\Str\sigma^2+\imath\Str\sigma\kappa\right)d[\sigma]&\propto&1-\frac{(\kappa_1-\kappa_2)}{2\pi\imath}\int\frac{\exp\left(-[s_1^2+s_2^2]/2+\imath s_1\kappa_1+s_2\kappa_2\right)}{s_1-\imath s_2}d[s].\nonumber\\
\fl&&\label{3.0}
\end{eqnarray}
The matrix $\sigma$ is a Hermitian $(1+1)\times(1+1)$ supermatrix with the bosonic eigenvalue $s_1$ and the fermionic eigenvalue $\imath s_2$ and $\kappa=\diag(\kappa_1,\kappa_2)$ is a diagonal $(1+1)\times(1+1)$ supermatrix. The first term on the right hand side of Eq.~\eref{3.0} is the Efetov-Wegner term and the second one is the supergroup integral appearing after diagonalizing $\sigma=UsU^\dagger$ with $U\in\U(1/1)$. Indeed, this result can be easily generalized to higher dimensional supermatrices with Gaussian weights. However, considering Gaussian weights is not sufficient to generalize this result to arbitrary weights. Thus, we pursue another approach.

The main idea of our approach is the comparison of results for generating functions,
\begin{equation}\label{3.1}
 Z_{k_1/k_2}^{(N)}(\kappa)=\int\limits_{\Herm(N)}P^{(N)}(H)\frac{\prod\limits_{j=1}^{k_2}\det(H-\kappa_{j2}\eins_{N})}{\prod\limits_{j=1}^{k_1}\det(H-\kappa_{j1}\eins_{N})}d[H]\,,
\end{equation}
obtained with and without supersymmetry. The matrix $\eins_N$ is the $N$ dimensional unit matrix. The integration domain in Eq.~\eref{3.1} is the set $\Herm(N)$ of $N\times N$ Hermitian matrices and the parameter $\kappa=\diag(\kappa_{11},\ldots,\kappa_{k_11},\kappa_{12},\ldots,\kappa_{k_22})=\diag(\kappa_1,\kappa_2)$ are chosen in such a way that the integral exists. One common choice of them is as real energies $x_j$ with a small imaginary part $\imath\varepsilon$ and source variables $J_j$, i.e. $\kappa_{j1/2}=x_j\mp J_j-\imath\varepsilon$ \cite{Guh06,KGG08}. The differentiation with respect to $J_j$ of $Z_{k_1/k_2}^{(N)}$ generates the matrix Green function which are intimately related to the $k$-point correlation functions \cite{Guh91}. The measure $d[H]$ is defined as
\begin{equation}\label{3.2}
 d[H]=\prod\limits_{n=1}^NdH_{nn}\prod\limits_{1\leq m< n\leq N}^Nd\RE\,H_{mn}d\IM\,H_{mn}\,.
\end{equation}
The matrix $\eins_N$ is the $N\times N$ unit matrix and $P^{(N)}$ is a probability density over $N\times N$ Hermitian matrices.

In Sec.~\ref{sec3}, we show that $Z_{k_1/k_2}^{(N)}$ is a determinant,
 \begin{equation}
  Z_{k_1/k_2}^{(N)}(\kappa)\sim\det\left[\begin{array}{c|c} \displaystyle\frac{Z_{1/1}^{(N)}(\kappa_{a1},\kappa_{b2})}{\kappa_{a1}-\kappa_{b2}} & \displaystyle Z_{1/0}^{(b)}(\kappa_{a1}) \end{array}\right]\,,\label{2.3}
 \end{equation}
if the Fourier-transform
\begin{equation}\label{3.3}
 \mathcal{F}P^{(N)}(\widetilde{H})=\int\limits_{\Herm(N)}P^{(N)}(H)\exp\left[\imath\tr H\widetilde{H}\right]d[H]
\end{equation}
factorizes in the eigen-representation of the Hermitian matrix $\widetilde{H}$. These determinantal structures are similar to those found for generating functions with factorizing $P^{(N)}$ in the eigen-representation of the Hermitian matrix $H$ \cite{BorStr06,KieGuh09a}.

Moreover, the integral of $Z_{k_1/k_2}^{(N)}$ can be easily mapped into superspace if the characteristic function $\mathcal{F}P^{(N)}$ is rotation invariant, see Refs.~\cite{Guh06,KGG08}. In this representation one does not integrate over ordinary matrices but over Wick-rotated Hermitian supermatrices \cite{KKG08,KGG08}. The integrand is almost rotation invariant apart from an exponential term which reflects a Fourier-transformation in superspace. We aim at the full integrand with all Efetov-Wegner terms appearing by a diagonalization of a supermatrix. In particular we want to find the distribution $\widehat{\varphi}_{k_1/k_2}$ which satisfies
\begin{equation}\label{2.4}
 \int\exp\left[-\imath\Str \kappa \rho\right]F(\rho)d[\rho]=\int\Ber_{k_1/k_2}^{(2)}(r)\widehat{\varphi}_{k_1/k_2}(-\imath r,\kappa)F(r)d[r]
\end{equation}
for an arbitrary rotation invariant superfunction $F$. In Eq.~\eref{2.4} we diagonalize the $(k_1+k_2)\times(k_1+k_2)$ Hermitian supermatrix $\rho$ to its eigenvalues $r$. This diagonalization does not only yield the Berezinian $\Ber_{k_1/k_2}^{(2)}$ but also the distribution $\widehat{\varphi}_{k_1/k_2}$. The distribution $\widehat{\varphi}_{k_1/k_2}$ is the integral over the supergroup $\U(k_1/k_2)$ and, additionally, all Efetov-Wegner terms. It is called the supermatrix Bessel function with Efetov-Wegner terms \cite{GuhKoh02b,KKG08}.

We derive $\widehat{\varphi}_{k_1/k_2}$ in two steps. In the first step we combine the mapping into superspace with the determinant~\eref{2.3} for factorizing $\mathcal{F}P^{(N)}$. This is a particular case of the identity Eq.~\eref{2.4} but it is sufficient to generalize its result to an arbitrary rotation invariant superfunction $F$ in the second step.

The procedure described above incorporates determinants derived in Ref.~\cite{BasFor94,KieGuh09a}. Without loss of generality let $p\geq q$. Then, these determinants are
\begin{eqnarray}
 \sqrt{\Ber_{p/q}^{(2)}(\kappa)}&=&\displaystyle\frac{\prod\limits_{1\leq a<b\leq p}(\kappa_{a1}-\kappa_{b1})\prod\limits_{1\leq a<b\leq q}(\kappa_{a2}-\kappa_{b2})}{\prod\limits_{a=1}^{p}\prod\limits_{b=1}^{q}(\kappa_{a1}-\kappa_{b2})}\label{2.0a}\\
 &=&(-1)^{p(p-1)/2}\det\left[\begin{array}{c} \left\{\displaystyle\frac{\kappa_{b1}^{p-q}\kappa_{a2}^{q-p}}{\kappa_{b1}-\kappa_{a2}}\right\}\underset{1\leq b\leq p}{\underset{1\leq a\leq q}{\ }} \\ \left\{\displaystyle\kappa_{b1}^{a-1}\right\}\underset{1\leq b\leq p}{\underset{1\leq a\leq p-q}{\ }}\end{array} \right]\label{2.0b}\\
 &=&(-1)^{p(p-1)/2}\det\left[\begin{array}{c} \left\{\displaystyle\frac{1}{\kappa_{b1}-\kappa_{a2}}\right\}\underset{1\leq b\leq p}{\underset{1\leq a\leq q}{\ }} \\ \left\{\displaystyle\kappa_{b1}^{a-1}\right\}\underset{1\leq b\leq p}{\underset{1\leq a\leq p-q}{\ }}\end{array} \right]\,.\label{2.0c}
\end{eqnarray}
All three expressions find their applications in our discussion. For $p=q$ we obtain the Cauchy-determinant \cite{Guh91}
\begin{eqnarray}
 \sqrt{\Ber_{p/p}^{(2)}(\kappa)}&=&(-1)^{p(p-1)/2}\det\left[\displaystyle\frac{1}{\kappa_{a1}-\kappa_{b2}} \right]_{1\leq a,b\leq p}\,,\label{2.1}
\end{eqnarray}
whereas for $q=0$ we have the Vandermonde-determinant
\begin{eqnarray}
 \sqrt{\Ber_{p/0}^{(2)}(\kappa)}&=&\Delta_p(\kappa_1)=(-1)^{p(p-1)/2}\det\left[ \displaystyle\kappa_{b1}^{a-1} \right]_{1\leq a,b\leq p}\,.\label{2.2}
\end{eqnarray}
These determinants appear as square roots of Berezinians by diagonalizing Hermitian supermatrices. This also explains our notation for them. The upper index $2$ results from the Dyson index $\beta$ which is two throughout this work. Thus we are consistent with our notation used in other articles~\cite{KGG08,KSG09,KieGuh09a,KieGuh09b}.

\section{The determinantal structure and the characteristic function}\label{sec3}

We consider the generating function~\eref{3.1}. Let the probability density $P^{(N)}$ be rotation invariant. Then, the characteristic function~\eref{3.3} inherits this symmetry. We consider such characteristic functions which factorize in the eigenvalues $\widetilde{E}=\diag(\widetilde{E}_1,\ldots,$ $\widetilde{E}_N)$ of the matrix $\widetilde{H}$, i.e.
\begin{equation}\label{3.4}
 \mathcal{F}P^{(N)}(\widetilde{E})=\prod\limits_{j=1}^Nf(\widetilde{E}_j)\,.
\end{equation}
Due to the normalization of $P^{(N)}$, the function $f: \mathbb{R}\to\mathbb{C}$ is unity at zero. We restrict us to the case
\begin{equation}\label{3.5}
 k_2\leq k_1\leq N\,.
\end{equation}
This case is sufficient for our purpose.

 Let $f$ be conveniently integrable and $d=N+k_2-k_1$. We derive in \ref{app1} that the generating function in Eq.~\eref{3.1} has for factorizing characteristic function~\eref{3.4} under the condition~\eref{3.5} the determinantal expression
 \begin{eqnarray}
  \fl Z_{k_1/k_2}^{(N)}(\kappa)&=&
  \frac{(-1)^{k_2(k_2+1)/2+k_2k_1}}{\sqrt{\Ber_{k_1/k_2}^{(2)}(\kappa)}}\det\left[\begin{array}{cc} \left\{\displaystyle\frac{Z_{1/1}^{(\widetilde{N})}(\kappa_{a1},\kappa_{b2})}{\kappa_{a1}-\kappa_{b2}}\right\}\underset{1\leq b\leq k_2}{\underset{1\leq a\leq k_1}{\ }} & \left\{\displaystyle Z_{1/0}^{(b)}(\kappa_{a1})\right\}\underset{d+1\leq b\leq N}{\underset{1\leq a\leq k_1}{\ }} \end{array}\right]\nonumber\\
 \fl&&\label{3.6}
 \end{eqnarray}
 for all $\widetilde{N}\in\{d,d+1,\ldots,N\}$. Notice that Eq.~\eref{3.6} is indeed true for all values $\widetilde{N}\in\{d,d+1,\ldots,N\}$ because the determinant is skew symmetric. Hence, we may add any linear combination of the last $k_1-k_2$ columns to the first $k_2$ columns.

With this theorem we made the first step to obtain all Efetov-Wegner terms for the diagonalization of a supermatrix. For non-normalized probability densities, we find Eq.~\eref{3.6} multiplied by $f^{\lambda}(0)$ where $\lambda=d(d+1)/2-(N-1)N/2-\widetilde{N}k_2$. We need the non-normalized version to analyze the integrals in Sec.~\ref{sec5}.

Equation~\eref{3.6} is also a very handy intermediate result. It yields a result of a simpler structure than the one for factorizing probability densities \cite{BorStr06,KieGuh09a}. Thus, all eigenvalue correlations for probability densities stemming from a characteristic function with the property~\eref{3.4} are determined by one and two point averages.

\section{Integral representation in the superspace of the generating function and the Efetov--Wegner term of $Z^{(N)}_{1/1}$}\label{sec4}

In Refs.~\cite{Guh06,LSZ07,KGG08,KSG09}, it was shown that the generating function~\eref{3.1} can be mapped to an integral over supermatrices. Let $\Sigma_{k_1/k_2}^{(0)}$ be the set of supermatrices with the form
\begin{equation}\label{4.1}
 \rho=\left[\begin{array}{cc} \rho_1 & \eta^\dagger \\ \eta & \rho_2 \end{array}\right]\,.
\end{equation}
The Boson--Boson block $\rho_1$ is an ordinary $k_1\times k_1$ positive definite Hermitian matrix and the Fermion--Fermion block $\rho_2$ is an ordinary $k_2\times k_2$ Hermitian matrix. The off-diagonal block $\eta$ comprises $k_1\times k_2$ independent Grassmann variables. We recall that $(\eta^\dagger)^\dagger=-\eta$ and the integration over one Grassmann variable is defined by
\begin{equation}\label{4.2}
 \int\eta_{nm}^jd\eta_{nm}=\int\eta_{nm}^{*\,j}d\eta_{nm}^*=\delta_{j1}\,,\quad j\in\{0,1\}\,.
\end{equation}
We need the Wick--rotated set $\Sigma_{k_1/k_2}^{(\psi)}=\widetilde{\Pi}_\psi\Sigma_{k_1/k_2}^{(0)}\widetilde{\Pi}_\psi$ to regularize the integrals below. The matrix $\widetilde{\Pi}_\psi=(\eins_{k_1},e^{\imath\psi/2}\eins_{k_2})$ with $\psi\in]0,\pi[$ is the generalized Wick--rotation \cite{KKG08,KGG08}. We assume that a Wick--rotation exists such that the characteristic function is a Schwartz function on the Wick--rotated real axis. Examples of such functions are given in Ref.~\cite{KGG08}.

We define the supersymmetric extension $\Phi$ of the characteristic function $\mathcal{F}P$ with help of a representation
\begin{equation}\label{4.2b}
 \mathcal{F}P_{1}(\tr H^m, m\in\mathbb{N})=\mathcal{F}P^{(N)}(H)
\end{equation}
as a function in matrix invariants,
\begin{equation}\label{4.2c}
 \Phi^{(k_1/k_2)}(\rho)=\mathcal{F}P_{1}(\Str\rho^m, m\in\mathbb{N})\,.
\end{equation}
For a factorizing characteristic function~\eref{3.4} we want to consider only such extensions $\Phi^{(k_1/k_2)}$ which factorize, too. In particular, we only use the extension
 \begin{equation}\label{4.11}
  \Phi^{(k_1/k_2)}(\rho)=\prod\limits_{a=1}^{k_1}f(r_{a1})\prod\limits_{b=1}^{k_2}\frac{1}{f(e^{\imath\psi}r_{b2})}\,.
 \end{equation}
Indeed the extension of $\mathcal{F}P^{(N)}$ to $\Phi^{(k_1/k_2)}$ is not unique and, hence, the factorization property~\eref{4.11} does not hold for each extension, see Ref.~\cite{KGG08}. However, for our purpose the choice~\eref{4.11} is sufficient at the moment. Later on we extend our result to arbitrary rotation invariant superfunctions such that it also applies to other rotation invariant extensions.

Let the signs of the imaginary parts for all $\kappa_{j1}$ be negative. Assuming that $\Phi$ is analytic in the Fermion--Fermion block $\rho_2$, the generalized Hubbard-Stratonovich transformation \cite{Guh06,KGG08,KSG09} tells us that the integral Eq.~\eref{3.1} with the condition \eref{3.5} is
\begin{eqnarray}
 \fl Z_{k_1/k_2}^{(N)}(\kappa)&=&C_{k_1/k_2}^{(N)}\int\limits_{\Sigma_{k_1/k_2}^{(\psi)}}\Phi^{(k_1/k_2)}(\hat{\rho})\exp[-\imath\Str\kappa\hat{\rho}]{\det}^d\rho_1\prod\limits_{j=1}^{k_2}\left(e^{-\imath\psi}\frac{\partial}{\partial r_{j2}}\right)^{d-1}e^{-\imath\psi}\delta\left(r_{j2}\right) d[\rho]\,,\nonumber\\
\fl&&\label{4.3}
\end{eqnarray}
where $e^{-\imath\psi}r_{j2}$ are the eigenvalues of $\rho_2$ and the matrix $\hat{\rho}$ is given by
\begin{equation}\label{4.3b}
 \hat{\rho}=\left[\begin{array}{c|c} \rho_1 & e^{\imath\psi/2}\eta^\dagger \\ \hline e^{\imath\psi/2}\eta & e^{\imath\psi}(\overset{\ }{\rho_2+\eta\rho_1^{-1}\eta^\dagger}) \end{array}\right]\,.
\end{equation}
The constant is
\begin{equation}\label{4.4}
 \fl C_{k_1/k_2}^{(N)}=\frac{(-1)^{k_2(k_2+2N-1)/2}\imath^{N(k_2-k_1)}\pi^{N(k_2-k_1)+k_2}2^{k_2k_1+k_2-k_1}}{\left[(d-1)!\right]^{k_2}}\frac{{\rm Vol}(\U(N))}{{\rm Vol}(\U(d))}f^d(0)\,.
\end{equation}
Here, the volume of the unitary group $\U(N)$ is
\begin{equation}\label{4.5}
 {\rm Vol}(\U(N))=\prod\limits_{j=1}^N\frac{2\pi^j}{(j-1)!}\,.
\end{equation}
The term $f(0)$ becomes unity if we consider normalized probability densities. The definition of the measure $d[\rho]=d[\rho_1]d[\rho_2]d[\eta]$ is equal to the one in Ref.~\cite{KSG09},
\begin{eqnarray}
 d[\rho_1]&=&\prod\limits_{n=1}^{k_1}d\rho_{nn1}\prod\limits_{1\leq m< n\leq k_1}d\RE\,\rho_{mn1}d\IM\,\rho_{mn1}\,,\label{4.6}\\
 d[\rho_2]&=&e^{\imath k_2^2\psi}\prod\limits_{n=1}^{k_2}d\rho_{nn2}\prod\limits_{1\leq m< n\leq k_2}d\RE\,\rho_{mn2}d\IM\,\rho_{mn2}\,,\label{4.7}\\
 d[\eta]&=&e^{-\imath k_1k_2\psi}\prod\limits_{n=1}^{k_1}\prod\limits_{m=1}^{k_2}d\eta_{mn}d\eta_{mn}^*\,.\label{4.8}
\end{eqnarray}
We use the conventional notation for the supertrace ``$\Str$'' and superdeterminant ``$\Sdet$''.

Let $\rho\in\Sigma_{k_1/k_2}^{(\psi)}$ with the form \eref{4.1}. As long as the eigenvalues of the Boson-Boson block $\rho_1$ are pairwise different with those of the Fermion-Fermion block $\rho_2$, we may diagonalize the whole supermatrix $\rho$ by an element $U\in \U(k_1/k_2)$. The corresponding diagonal eigenvalue matrix is $r=\diag(r_{11},\ldots,r_{k_11},e^{\imath\psi}r_{12},\ldots,e^{\imath\psi}r_{22})=\diag(r_1,e^{\imath\psi}r_2)$, i.e. $\rho=UrU^\dagger$. Due to Rothstein's \cite{Rot87} vector field resulting from such a change of coordinates in the Berezin measure, we have
\begin{equation}\label{4.10}
 d[\rho]\neq\Ber_{k_1/k_2}^{(2)}(r)d[r]d\mu(U)\,,
\end{equation}
where $d[r]$ is the product of all eigenvalue differentials and $d\mu(U)$ is the supersymmetric Haar--measure of the unitary supergroup $\U(k_1/k_2)$. We have to consider some boundary terms since the Berezin integral is fundamentally connected with differential operators \cite{Rot87,BieSom07,KKG08,BES09}.

We consider the generating function $Z_{1/1}^{(N)}$. Equation~\eref{4.3} is an integral over Dirac distributions. Hence, we cannot simply apply a Cauchy-like theorem \cite{Weg83,Efe83,Con88,ConGro89,Efe97,KKG08} but also for this integration domain we obtain an Efetov--Wegner term for $Z_{1/1}^{(N)}$.
 Let the function $1/f$ be analytic at zero. Then, we have
 \begin{eqnarray}
  \fl\frac{Z_{1/1}^{(N)}(\kappa)}{f^N(0)}&=&\left.\left[\frac{f(r_1)}{f\left(e^{\imath\psi}r_2\right)}\exp[-\imath\Str\kappa r]\right]\right|_{r=0}+\frac{\imath(-1)^{N}}{(N-1)!}\nonumber\\
  \fl&\times&\int\limits_{\mathbb{R}_+\times\mathbb{R}}\frac{\kappa_1-\kappa_2}{r_1-e^{\imath\psi}r_2}\frac{f(r_1)}{f\left(e^{\imath\psi}r_2\right)}\exp[-\imath\Str\kappa r]r_1^N\left(e^{-\imath\psi}\frac{\partial}{\partial r_{2}}\right)^{N-1}\delta\left(r_{2}\right)dr_2dr_1\label{4.12}\\
 \fl&=&\frac{(-1)^{N}}{(N-1)!}\int\limits_{\mathbb{R}_+\times\mathbb{R}}\left[\frac{1}{r_1-e^{\imath\psi}r_2}\left(\frac{\partial}{\partial r_1}+e^{-\imath\psi}\frac{\partial}{\partial r_2}\right)+\imath\frac{\kappa_1-\kappa_2}{r_1-e^{\imath\psi}r_2}\right]\nonumber\\
 \fl&\times&\frac{f(r_1)}{f\left(e^{\imath\psi}r_2\right)}\exp[-\imath\Str\kappa r]r_1^N\left(e^{-\imath\psi}\frac{\partial}{\partial r_{2}}\right)^{N-1}\delta\left(r_{2}\right)dr_2dr_1\,.\label{4.13}
 \end{eqnarray}
 In Eq.~\eref{4.13}, we, first, integrate over $r_2$ and, then, over $r_1$. We derive this result in \ref{app2}. The first summand on the right hand side of equality~\eref{4.12} is $1$ which is the Efetov-Wegner term. The second equality~\eref{4.13} is more convenient than equality~\eref{4.12} for the discussions in the ensuing section.

\section{Supermatrix Bessel function with all Efetov--Wegner terms}\label{sec5}

Let supermatrices in $\widetilde{\Sigma}_{k_1/k_2}^{(\psi)}$ be similar to those in $\Sigma_{k_1/k_2}^{(\psi)}$ without the positive definiteness of the Boson-Boson block. We want to find the distribution $\widehat{\varphi}_{k_1/k_2}$ which satisfies
\begin{equation}\label{5.1}
 \fl\int\limits_{\widetilde{\Sigma}_{k_1/k_2}^{(\psi)}}\exp\left[-\imath\Str \kappa \rho\right]F(\rho)d[\rho]=\int\limits_{\mathbb{R}^{k_1+k_2}}\Ber_{k_1/k_2}^{(2)}(r)\widehat{\varphi}_{k_1/k_2}(-\imath r,\kappa)F(r)d[r]\,,
\end{equation}
for an arbitrary sufficiently integrable, rotation invariant superfunction $F$ analytic at zero. Recognizing that the integral expression~\eref{4.3} includes the supersymmetric Ingham--Siegel integral \cite{Guh06,KGG08,KSG09}, the generating function is apart from a shift in the Fermion-Fermion block and analyticity a particular example of this type of integral.

First we want to derive $\widehat{\varphi}_{k_1/k_2}$ in Eq.~\eref{5.1} for factorizing superfunctions~\eref{4.11} and, then, extend to arbitrary $F$. We show in \ref{app3} that the distribution $\widehat{\varphi}_{k_1/k_2}$ defined by
 \begin{eqnarray}
  \fl&&\int\limits_{\Sigma_{k_1/k_2}^{(\psi)}}F(\hat{\rho})\exp[-\imath\Str\kappa\hat{\rho}]{\det}^d\rho_1\prod\limits_{j=1}^{k_2}\left(e^{-\imath\psi}\frac{\partial}{\partial r_{j2}}\right)^{d-1}e^{-\imath\psi}\delta\left(r_{j2}\right)d[\rho]\nonumber\\
  \fl&=&\int\limits_{\mathbb{R}_+^{k_1}\times\mathbb{R}^{k_2}}\Ber_{k_1/k_2}^{(2)}(r)\widehat{\varphi}_{k_1/k_2}(-\imath r,\kappa)F(r){\det}^d r_1\prod\limits_{j=1}^{k_2}\left(e^{-\imath\psi}\frac{\partial}{\partial r_{j2}}\right)^{d-1}e^{-\imath\psi}\delta\left(r_{j2}\right)d[r]\label{5.5}
 \end{eqnarray}
 is given by
 \begin{eqnarray}
  \fl\widehat{\varphi}_{k_1/k_2}(-\imath r,\kappa)&=&\frac{(-1)^{(k_1+k_2)(k_1+k_2-1)/2}(\imath\pi)^{(k_2-k_1)^2/2-(k_1+k_2)/2}}{2^{k_1k_2}k_1!k_2!\sqrt{\Ber_{k_1/k_2}^{(2)}(\kappa)}\Ber_{k_1/k_2}^{(2)}(r)}\nonumber\\
 \fl&\times&\underset{\omega_2\in\mathfrak{S}(k_2)}{\underset{\omega_1\in\mathfrak{S}(k_1)}{\sum}}\exp\left[-\imath\sum\limits_{j=1}^{k_1}\kappa_{j1}r_{\omega_1(j)1}+\imath e^{\imath\psi}\sum\limits_{j=1}^{k_2}\kappa_{j2}r_{\omega_2(j)2}\right]\label{5.6}\\
  \fl&\times&\det\left[\begin{array}{c} \left\{\displaystyle \frac{-\imath}{(\kappa_{b1}-\kappa_{a2})(r_{\omega_1(b)1}-e^{\imath\psi}r_{\omega_2(a)2})}\left(\frac{\partial}{\partial r_{\omega_1(b)1}}+e^{-\imath\psi}\frac{\partial}{\partial r_{\omega_2(a)2}}\right)\right\}\underset{1\leq b\leq k_1}{\underset{1\leq a\leq k_2}{\ }} \\ \left\{\displaystyle r_{\omega_1(b)1}^{a-1}\right\}\underset{1\leq b\leq k_1}{\underset{1\leq a\leq k_1-k_2}{\ }} \end{array}\right]\nonumber
 \end{eqnarray}
 for an arbitrary sufficiently integrable, rotation invariant superfunction $F$ analytic at zero and factorizing in the eigen representation of $\rho$. The set of permutations over $k$ elements is $\mathfrak{S}(k)$.  The derivation of Eq.~\eref{5.6} incorporates Eqs.~\eref{3.6} and \eref{4.13}.

In \ref{app4} we make the next step and generalize Eq.~\eref{5.6} to arbitrary rotation invariant superfunctions without the additional Dirac-distributions in the integral, cf. Eq.~\eref{4.3}. The distribution $\widehat{\varphi}_{k_1/k_2}$ defined by Eq.~\eref{5.1} is indeed the one defined in Eq.~\eref{5.6} for an arbitrary sufficiently integrable, rotation invariant superfunction $F(\rho)$ analytic at zero. For such superfunctions and for generalized Wick-rotations $\psi\in]0,\pi[$ this distribution has alternatively the form
 \begin{eqnarray}
  \fl&&\widehat{\varphi}_{k_1/k_2}(-\imath r,\kappa)
  =\frac{(-1)^{(k_1+k_2)(k_1+k_2-1)/2}(\imath\pi)^{(k_2-k_1)^2/2-(k_1+k_2)/2}}{2^{k_1k_2}k_1!k_2!\sqrt{\Ber_{k_1/k_2}^{(2)}(\kappa)}\Ber_{k_1/k_2}^{(2)}(r)}\underset{\omega_2\in\mathfrak{S}(k_2)}{\underset{\omega_1\in\mathfrak{S}(k_1)}{\sum}}\label{5.8}\\
  \fl&\times&\det\left[\begin{array}{c} \left\{\displaystyle \frac{-2\pi e^{-\imath\psi}\delta(r_{\omega_1(b)1})\delta(r_{\omega_2(a)2})}{\kappa_{b1}-\kappa_{a2}}+\frac{\exp\left(-\imath\kappa_{b1}r_{\omega_1(b)1}+\imath e^{\imath\psi}\kappa_{a2}r_{\omega_2(a)2}\right)}{r_{\omega_1(b)1}-e^{\imath\psi}r_{\omega_2(a)2}}\chi(\kappa_{b1}-\kappa_{a2})\right\}\underset{1\leq b\leq k_1}{\underset{1\leq a\leq k_2}{\ }} \\ \left\{\displaystyle r_{\omega_1(b)1}^{a-1}\exp\left(-\imath\kappa_{b1}r_{\omega_1(b)1}\right)\right\}\underset{1\leq b\leq k_1}{\underset{1\leq a\leq k_1-k_2}{\ }} \end{array}\right]\nonumber
 \end{eqnarray}
 with the distribution
 \begin{equation}\label{5.10}
  \chi(x)=\left\{\begin{array}{cl} 0 &,\ x=0,\\ 1 &,\ \mathrm{else.}\end{array} \right.
 \end{equation}
Equation~\eref{5.8} is true because of the Cauchy-like theorem for $(1+1)\times(1+1)$ Hermitian supermatrices, see Refs.~\cite{Weg83,ConGro89,KKG08}. One has to careful on which half of the complex plane the general Wick-rotation is lying. If $\psi\in]\pi,2\pi[$ then the minus sign changes to a plus infront of the Dirac-distributions.

 We emphasize that the distribution $\widehat{\varphi}_{k_1/k_2}(r,\kappa)$ is not symmetric in exchanging its arguments $r$ and $\kappa$. Apart from the characteristic function $\chi$ such a symmetry exists for the supermatrix Bessel functions \cite{Guh91,Guh96,KKG08} which is $\widehat{\varphi}_{k_1/k_2}(r,\kappa)$ without the Dirac-distributions, i.e.
\begin{eqnarray}\label{5.9}
 \fl\varphi_{k_1/k_2}(-\imath r,\kappa)&=&\frac{(-1)^{(k_1+k_2)(k_1+k_2-1)/2}(\imath\pi)^{(k_2-k_1)^2/2-(k_1+k_2)/2}}{2^{k_1k_2}k_1!k_2!\sqrt{\Ber_{k_1/k_2}^{(2)}(\kappa)}\Ber_{k_1/k_2}^{(2)}(r)}\\
\fl&\times&\underset{\omega_2\in\mathfrak{S}(k_2)}{\underset{\omega_1\in\mathfrak{S}(k_1)}{\sum}}\det\left[\begin{array}{c} \left\{\displaystyle \frac{\exp\left(-\imath\kappa_{b1}r_{\omega_1(b)1}+\imath e^{\imath\psi}\kappa_{a2}r_{\omega_2(a)2}\right)}{r_{\omega_1(b)1}-e^{\imath\psi}r_{\omega_2(a)2}}\chi(\kappa_{b1}-\kappa_{a2})\right\}\underset{1\leq b\leq k_1}{\underset{1\leq a\leq k_2}{\ }} \\ \left\{\displaystyle r_{\omega_1(b)1}^{a-1}\exp\left(-\imath\kappa_{b1}r_{\omega_1(b)1}\right)\right\}\underset{1\leq b\leq k_1}{\underset{1\leq a\leq k_1-k_2}{\ }} \end{array}\right]\nonumber\\
 \fl&=&\frac{(-1)^{k_2(k_2-1)/2+k_1k_2}(\imath\pi)^{(k_2-k_1)^2/2-(k_1+k_2)/2}}{2^{k_1k_2}k_1!k_2!}\underset{1\leq b\leq k_2}{\underset{1\leq a\leq k_1}{\prod}}\chi(\kappa_{a1}-\kappa_{b2})\nonumber\\
 \fl&\times&\frac{\det\left[\exp(-\imath\kappa_{a1}r_{b1})\right]_{1\leq a,b\leq k_1}\det\left[\exp(\imath e^{\imath\psi}\kappa_{a2}r_{b2})\right]_{1\leq a,b\leq k_2}}{\sqrt{\Ber_{k_1/k_2}^{(2)}(\kappa)}\sqrt{\Ber_{k_1/k_2}^{(2)}(r)}}\nonumber\,.
\end{eqnarray}
The asymmetry, $\widehat{\varphi}_{k_1/k_2}(r,\kappa)\neq\widehat{\varphi}_{k_1/k_2}(\kappa,r)$, is mainly due to the diagonalization of $\rho$ to $r$ whereas the supermatrix $\kappa$ is already diagonal. The characteristic function $\chi$ in Eqs.~\eref{5.8} and \eref{5.9} is crucial because of the commutator
\begin{eqnarray}
 &&\left[\frac{\partial}{\partial r_{\omega_1(b)1}}+e^{-\imath\psi}\frac{\partial}{\partial r_{\omega_2(a)2}},\exp\left(-\imath\kappa_{b1}r_{\omega_1(b)1}+\imath e^{\imath\psi}\kappa_{a2}r_{\omega_2(a)2}\right)\right]_-\nonumber\\
 &=&-\imath(\kappa_{b1}-\kappa_{a2})\exp\left(-\imath\kappa_{b1}r_{\omega_1(b)1}+\imath e^{\imath\psi}\kappa_{a2}r_{\omega_2(a)2}\right)\chi(\kappa_{b1}-\kappa_{a2})\,.\label{5.11}
\end{eqnarray}
Indeed the set which is cutted out by $\chi$ is a set of measure zero and does not play any role when one integrates $\varphi_{k_1/k_2}(-\imath r,\kappa)$ or $\widehat{\varphi}_{k_1/k_2}(-\imath r,\kappa)$ over $\kappa$ with a function continuous around this set. However it becomes important for the integral
\begin{eqnarray}\label{5.12}
 \fl&&\int\limits_{\mathbb{R}^{k_1+k_2}}\varphi_{k_1/k_2}(\imath s,r)\varphi_{k_1/k_2}(-\imath r,\widetilde{\kappa})\Ber^{(2)}_{k_1/k_2}(r)d[r]\\
 \fl&=&\frac{\pi^{(k_2-k_1)^2}}{2^{2k_1k_2-k_1-k_2}k_1!k_2!}\underset{1\leq b\leq k_2}{\underset{1\leq a\leq k_1}{\prod}}\chi(\kappa_{a1}-e^{-\imath\psi}\kappa_{b2})\frac{\det\left[\delta(\kappa_{a1}-s_{b1})\right]_{1\leq a,b\leq k_1}\det\left[e^{\imath\psi}\delta(\kappa_{a2}-s_{b2})\right]_{1\leq a,b\leq k_2}}{\sqrt{\Ber_{k_1/k_2}^{(2)}(\widetilde{\kappa})}\sqrt{\Ber_{k_1/k_2}^{(2)}(s)}}\nonumber\,.
\end{eqnarray}
where $s=\diag(s_{11},\ldots,s_{k_11},e^{-\imath\psi} s_{12},\ldots,e^{-\imath\psi}s_{k_22})$ and $\widetilde{\kappa}=\diag(\kappa_{11},\ldots,\kappa_{k_11},$ $e^{-\imath\psi}\kappa_{12},\ldots,e^{-\imath\psi}\kappa_{k_22})$ with $s_{ab},\kappa_{ab}\in\mathbb{R}$. This result is the correct one for the supermatrix Bessel function. The difference to other results \cite{Guh91,Guh96,KKG08} is the distribution $\chi$ which guarantees that the Dirac-distribution~\eref{5.12} in the eigenvalues $s$ of a Hermitian supermatrix $\sigma$ vanishes if a bosonic eigenvalue of $\widetilde{\kappa}$ equals to a fermionic one. On the first sight this would also happen if we omit $\chi$ due to the prefactor $\sqrt{\Ber_{k_1/k_2}^{(2)}(\widetilde{\kappa})}$ in Eq.~\eref{5.12} which vanishes on this set. Yet after applying the distribution~\eref{5.12} on functions over the set of diagonalized Hermitian supermatrices comprising the Berezinian $\Ber_{k_1/k_2}^{(2)}(s)$ in the measure this term cancels out, i.e.
\begin{eqnarray}
 \fl F(\kappa)&=&\int\limits_{\mathbb{R}^{k_1+k_2}}F(s)\frac{\det\left[\delta(\kappa_{a1}-s_{b1})\right]_{1\leq a,b\leq k_1}\det\left[e^{\imath\psi}\delta(\kappa_{a2}-s_{b2})\right]_{1\leq a,b\leq k_2}}{k_1!k_2!\sqrt{\Ber_{k_1/k_2}^{(2)}(\widetilde{\kappa})}\sqrt{\Ber_{k_1/k_2}^{(2)}(s)}}\Ber_{k_1/k_2}^{(2)}(s)d[s]\nonumber\\
 \fl&\neq&F(\kappa)\underset{1\leq b\leq k_2}{\underset{1\leq a\leq k_1}{\prod}}\chi(\kappa_{a1}-e^{-\imath\psi}\kappa_{b2})\,.\label{5.12b}
\end{eqnarray}
The reasoning why Eq.~\eref{5.12} has to vanish on this set becomes clear when we interpret Eq.~\eref{5.12} as an integral over the supergroup $\U(k_1/k_2)$, i.e.
\begin{eqnarray}\label{5.13}
 \fl\int\limits_{\mathbb{R}^{k_1+k_2}}\varphi_{k_1/k_2}(\imath s,r)\varphi_{k_1/k_2}(-\imath r,\widetilde{\kappa})\Ber^{(2)}_{k_1/k_2}(r)d[r]\sim\int\limits_{\U(k_1/k_2)}\delta(UsU^\dagger-\widetilde{\kappa})d\mu(U)\,.
\end{eqnarray}
The measure $d\mu(U)$ is the Haar measure on $\U(k_1/k_2)$ and the Dirac-distribution is defined by two Fourier transformations
\begin{eqnarray}\label{5.14}
 \delta(UsU^\dagger-\widetilde{\kappa})\sim\int\exp\left[\imath\Str\rho(UsU^\dagger-\widetilde{\kappa})\right]d[\rho]\,.
\end{eqnarray}
The Haar measure $d\mu$ of the supergroup $\U(k_1/k_2)$ cannot be normalized as it can be done for the ordinary unitary groups since the volume of $\U(k_1/k_2)$ is zero for $k_1k_2\neq0$. This is also the reason why Eq.~\eref{5.12} has to vanish if one bosonic eigenvalue of $\widetilde{\kappa}$ equals to a fermionic one. Then the integral~\eref{5.13} is rotation invariant under the subgroup $\U(1/1)$ which has zero volume, too. This cannot be achieved without the distribution $\chi$ as it was done in the common literature~\cite{Guh91,Guh96,KKG08}. Interestingly the replacement of $\varphi_{k_1/k_2}$ by $\widehat{\varphi}_{k_1/k_2}$ in Eq.~\eref{5.12} yields the full Dirac-distribution
\begin{eqnarray}\label{5.15}
 &&\int\limits_{\mathbb{R}^{k_1+k_2}}\widehat{\varphi}_{k_1/k_2}(\imath s,r)\widehat{\varphi}_{k_1/k_2}(-\imath r,\widetilde{\kappa})\Ber^{(2)}_{k_1/k_2}(r)d[r]\\
&=&\frac{\pi^{(k_2-k_1)^2}}{2^{2k_1k_2-k_1-k_2}k_1!k_2!}\frac{\det\left[\delta(\kappa_{a1}-s_{b1})\right]_{1\leq a,b\leq k_1}\det\left[e^{\imath\psi}\delta(\kappa_{a2}-s_{b2})\right]_{1\leq a,b\leq k_2}}{\sqrt{\Ber_{k_1/k_2}^{(2)}(\widetilde{\kappa})}\sqrt{\Ber_{k_1/k_2}^{(2)}(s)}}\nonumber
\end{eqnarray}
which is similar to the one in Eq.~\eref{5.12}. Equation~\eref{5.15} is derived in \ref{app5}.

We want to finish this section by a remark about the relation of the result for the supermatrix Bessel function, see Eq.~\eref{5.8}, and the differential operator derived by the author in an earlier work~\cite{KKG08}. This differential operator is defined by
\begin{equation}\label{5.16}
 D^{(k_1k_2)}_{r}F(r)=\int F(\rho)d[\eta]
\end{equation}
for an arbitrary sufficiently integrable superfunction $F$ on the $(k_1+k_2)\times(k_1+k_2)$ supermatrices invariant under $\U(k_1/k_2)$. It has the form
\begin{eqnarray}
 \fl D^{(k_1k_2)}_{r}&=&\frac{1}{(k_1k_2)!(4\pi)^{k_1k_2}}\frac{1}{\Delta_{k_1}(r_1)\Delta_{k_1}(e^{\imath\psi}r_2)}\label{5.17}\\
\fl&\times&\sum\limits_{n=0}^{k_1k_2}\left(\begin{array}{c} k_1k_2 \\ n\end{array}\right)\left(\Str\frac{\partial^2}{\partial r^2}\right)^{k_1k_2-n}\underset{1\leq b\leq k_2}{\underset{1\leq a\leq k_1}{\prod}}(r_{a1}-e^{\imath\psi}r_{b2})\left(-\Str\frac{\partial^2}{\partial r^2}\right)^n\sqrt{\Ber_{k_1/k_2}^{(2)}(r)}\,,\nonumber
\end{eqnarray}
where we define
\begin{equation}\label{5.18}
 \Str\frac{\partial^2}{\partial r^2}=\sum\limits_{a=1}^{k_1}\frac{\partial^2}{\partial r_{a1}^2}-e^{-2\imath \psi}\sum\limits_{b=1}^{k_2}\frac{\partial^2}{\partial r_{b2}^2}\,.
\end{equation}
Due to Eq.~\eref{5.16} the differential operator $D^{(k_1k_2)}_{r}$ is equivalent to the integration over all Grassmann variables of the supermatrix $\rho$.

The comparison of Eqs.~\eref{5.16} and \eref{5.17} with Eqs.~\eref{5.1} and \eref{5.8} for an arbitrary sufficiently integrable, rotation invariant superfunction $F$ and arbitrary diagonal supermatrix $\kappa$ yields
\begin{eqnarray}
\fl D^{(k_1k_2)}_{r}&=&\frac{1}{(2\pi)^{k_1k_2}}\frac{1}{\Delta_{k_1}(r_1)\Delta_{k_1}(e^{\imath\psi}r_2)}\underset{1\leq b\leq k_2}{\underset{1\leq a\leq k_1}{\prod}}\left(\frac{\partial}{\partial r_{a1}}+e^{-\imath\psi}\frac{\partial}{\partial r_{b2}}\right)\sqrt{\Ber_{k_1/k_2}^{(2)}(r)}\,.\label{5.19}
\end{eqnarray}
Thus, we have found a quite compact form for $D^{(k_1k_2)}_{r}$ which is easier to deal with as the one in Eq.~\eref{5.17}.

\section{Some applications for Hermitian matrix ensembles}\label{sec6}

In random matrix theory generating functions as
\begin{equation}\label{6.1}
 Z_{k}^{(N)}(\kappa, \alpha H_0)=\int\limits_{\Herm(N)}P^{(N)}(H)\prod\limits_{j=1}^{k}\frac{\det(H+\alpha H_0-\kappa_{j2}\eins_{N})}{\det(H+ \alpha H_0-\kappa_{j1}\eins_{N})}d[H]\,.
\end{equation}
are paramount important since they model Hermitian random matrices in external potentials \cite{BreHik98,Guh06b} or intermediate random matrix ensembles \cite{PanMeh83,Guh96b,Guh96c,Sto02,GuhSto04,Joh07}. The matrix $H_0$ is a $N\times N$ Hermitian matrix and can be arbitrarily chosen or can also be drawn from another random matrix ensemble. The external parameter $\alpha\in\mathbb{R}$ is the coupling constant between the two matrices $H$ and $H_0$ and yields the generating function~\eref{3.1} for $\alpha=0$, i.e. $Z_{k}^{(N)}(\kappa,0)=Z_{k/k}^{(N)}(\kappa)$. The variables $\kappa_{j1}$ have to have an imaginary increment to guarantee the convergence of the integral, i.e.  $\kappa_{j1}=x_{j1}-J_{j}-\imath\varepsilon$.

In subsection~\ref{sec6.1} we consider the generating function~\eref{6.1} with $\lambda=0$. We will use the mapping of this integral to a representation in superspace which is shown in a previous work by the authors \cite{KieGuh09a} and diagonalize the supermatrix. In a formalism similar to the case $\alpha=0$ we will treat the more general case $\alpha\neq0$ in subsection~\ref{sec6.2}.

\subsection{Hermitian matrix ensembles without an external field ($\alpha=0$)}\label{sec6.1}

Omitting the index $N$ in $P^{(N)}$ we consider a normalized rotation invariant probability density $P$ with respect to $N\times N$ Hermitian random matrices.

Following the derivations made in Refs.~\cite{Guh06,KGG08} we have to calculate the characteristic function $\mathcal{F}P$, see Eq.~\eref{3.3}. Assuming that this can be done we recall the rotation invariance of $\mathcal{F}P$. This allows us to choose a representation of $\mathcal{F}P$ as a function in a finite number of matrix invariants, i.e. $\mathcal{F}P(\widetilde{H})=\mathcal{F}P_1(\tr \widetilde{H},\ldots,\tr \widetilde{H}^N)$. A straightforward supersymmetric extension $\Phi$ of the characteristic function is the one in Eq.~\eref{4.2b} and its Fourier transform is
\begin{equation}\label{6.1.2}
 \mathcal{F}\Phi(\sigma)=2^{2k(k-1)}\int\limits_{\widetilde{\Sigma}_{k/k}^{(\psi)}}\Phi(\rho)\exp(-\imath\Str\rho\sigma)d[\rho]\,.
\end{equation}
The Fourier-transform is denoted by $Q$ in Ref.~\cite{Guh06}. The supermatrices $\rho$ and $\sigma$ are Wick-rotated with the phases $e^{\imath\psi}$ and $e^{-\imath\psi}$, respectively.

The supersymmetric integral for $Z_{k}^{(N)}(\kappa,\alpha H_0)$ is \cite{Guh06,KGG08}
\begin{eqnarray}
 Z_{k}^{(N)}(\kappa, \alpha H_0)&=&\int\limits_{\widetilde{\Sigma}_{k/k}^{(-\psi)}}\mathcal{F}\Phi(\sigma){\Sdet}^{-1}(\sigma\otimes\eins_N+\alpha\eins_{k+k}\otimes H_0-\kappa)d[\sigma]\,.\label{6.1.3}
\end{eqnarray}
Setting the coupling constant $\alpha$ to zero we also find
\begin{eqnarray}
 Z_{k}^{(N)}(\kappa)&=&2^{2k(k-1)}\int\limits_{\widetilde{\Sigma}_{k/k}^{(\psi)}}\int\limits_{\widetilde{\Sigma}_{k/k}^{(-\psi)}}\mathcal{F}\Phi(\sigma)\exp[\imath\Str\rho(\sigma-\kappa)]I_N(\rho)d[\sigma]d[\rho]\,,\label{6.1.4}
\end{eqnarray}
where the supersymmetric Ingham-Siegel integral is \cite{Guh06}
\begin{eqnarray}
 \fl I_k^{(N)}(\rho)&=&\int\limits_{\widetilde{\Sigma}_{k/k}^{(-\psi)}}{\Sdet}^{-N}(\sigma+\imath\varepsilon\eins_{k+k})\exp(-\imath\Str\rho\sigma+\varepsilon\Str\rho)d[\sigma]\nonumber\\
 \fl&=&(-1)^{k(k+1)/2}2^{-k(k-1)}\prod\limits_{j=1}^k\left[\frac{2\pi e^{-\imath\psi}}{(N-1)!}r_{j1}^N\Theta(r_{j1})\left(-e^{-\imath\psi}\frac{\partial}{\partial r_{j2}}\right)^{N-1}\delta(r_{j2})\right]\,,\label{6.1.5}
\end{eqnarray}
The distribution $\Theta$ is the Heavyside distribution and $r_{j1}$ and $e^{\imath\psi}r_{j2}$ are the bosonic and fermionic eigenvalues of $\rho$, respectively. In \ref{app6} we perform the integration~\eref{6.1.5} with help of the results in Sec.~\ref{sec5} and find
\begin{eqnarray}\label{6.1.6}
 \fl Z_{k}^{(N)}(\kappa)&=&\frac{\imath^k}{2^{k(k-1)}}\int\limits_{\mathbb{R}^{2k}}\frac{d[s]}{\sqrt{\Ber^{(2)}_{k/k}(\kappa)}}\mathcal{F}\Phi(s)\\
 \fl&\times&\hspace*{-0.05cm}\det\hspace*{-0.05cm}\left[\displaystyle\hspace*{-0.05cm} \frac{e^{\imath\psi}\delta(s_{b1})\delta(s_{a2})}{\kappa_{b1}-\kappa_{a2}}\left(\frac{\kappa_{a2}}{\kappa_{b1}}\right)^N\hspace*{-0.1cm}+\frac{N\left(e^{-\imath\psi}s_{a2}-\kappa_{a2}\right)^{N-1}\chi(\kappa_{b1}-\kappa_{a2})}{2\pi\imath\left(e^{-\imath\psi}s_{b1}-\kappa_{b1}\right)^{N+1}(s_{b1}-e^{-\imath\psi}s_{a2})}\right]_{1\leq a,b\leq k}\hspace*{-0.5cm}\,.\nonumber
\end{eqnarray}
The first term in the determinant is the Efetov-Wegner term whereas the second term can be understood as integrals over supergoups.

\subsection{Hermitian matrix ensembles in the presence of an external field ($\alpha\neq0$)}\label{sec6.2}

For Hermitian matrix ensembles in an external field it is convenient to consider the integral representation
\begin{eqnarray}
 \fl && Z_{k}^{(N)}(\kappa, \alpha H_0)=2^{2k(k-1)}\int\limits_{\widetilde{\Sigma}_{k/k}^{(\psi)}}\Phi(\rho)\exp[-\imath\Str\kappa\rho]\label{6.2.1}\\
 \fl &\times&\left[\int\limits_{\widetilde{\Sigma}_{k/k}^{(-\psi)}}\exp(-\imath\Str\rho\sigma+\varepsilon\Str\rho){\Sdet}^{-1}(\sigma\otimes\eins_N+\alpha\eins_{k+k}\otimes H_0+\imath\varepsilon\eins_{N(k+k)})d[\sigma]\right]d[\rho]\nonumber
\end{eqnarray}
for the generating function, see Eqs.~\eref{6.1} and \eref{6.1.3}. In \ref{app7} we integrate this representation in two steps and get
\begin{eqnarray}
 \fl Z_{k}^{(N)}(\kappa, \alpha H_0)&=&\frac{(-1)^{k(k-1)/2}}{\Delta_N(\alpha E_0)\sqrt{\Ber_{k/k}^{(2)}(\kappa)}}\label{6.2.2}\\
 \fl&&\int\limits_{\mathbb{R}^{2k}}\det\left[\begin{array}{cc}\displaystyle\left\{B_1(r_{b1},r_{a2},\kappa_{b1},\kappa_{a2})\right\}_{1\leq a,b\leq k} & \displaystyle\left\{B_{b2}(r_{a2},\kappa_{a2})\right\}\underset{1\leq b\leq N}{\underset{1\leq a\leq k}{\ }} \\  \displaystyle\left\{B_3(r_{b1},\kappa_{b1}, \alpha E_{a}^{(0)})\right\}\underset{1\leq b\leq k}{\underset{1\leq a\leq N}{\ }} & \displaystyle\left\{(-\alpha E_{a}^{(0)})^{b-1}\right\}_{1\leq a,b\leq N} \end{array}\right]\Phi(r)d[r]\,,\nonumber
\end{eqnarray}
where
\begin{eqnarray}
 \fl B_1(r_{b1},r_{a2},\kappa_{b1},\kappa_{a2})&=&\frac{e^{-\imath\psi}\delta(r_{b1})\delta(r_{a2})}{\kappa_{b1}-\kappa_{a2}}\,,\nonumber\\
 \fl&+&\imath\frac{I_1^{(N)}(r_{b1},e^{\imath\psi}r_{a2})}{2\pi(r_{b1}-e^{\imath\psi}r_{a2})}\exp\left(-\imath\kappa_{b1}r_{b1}+\imath e^{\imath\psi} \kappa_{a2}r_{a2}\right)\chi(\kappa_{b1}-\kappa_{a2})\label{6.2.3}\\
 \fl B_{b2}(r_{a2},\kappa_{a2})&=&\exp\left(\imath e^{\imath\psi} \kappa_{a2}r_{a2}\right)\left(-\imath e^{-\imath\psi}\frac{\partial}{\partial r_{a2}}\right)^{b-1}e^{-\imath\psi}\delta(r_{a2})\prod\limits_{j=1}^k\chi(\kappa_{j1}-\kappa_{a2})\,,\label{6.2.4}\\
 \fl B_3(r_{b1},\kappa_{b1}, \alpha E_{a}^{(0)})&=&-\imath\exp\left(-\imath\kappa_{b1}r_{b1}\right)\Theta(r_{b1})\sum\limits_{n=N}^\infty\frac{(\imath \alpha E_{a}^{(0)} r_{b1})^n}{n!}\prod\limits_{j=1}^k\chi(\kappa_{b1}-e^{\imath\psi}\kappa_{j2})\,.\label{6.2.5}
\end{eqnarray}
The case $\alpha=0$ can be easily deduced from the result~\eref{6.2.2},
\begin{eqnarray}
 \fl Z_{k}^{(N)}(\kappa)&=&\frac{(-1)^{k(k-1)/2}}{\sqrt{\Ber_{k/k}^{(2)}(\kappa)}}\int\limits_{\mathbb{R}^{2k}}\det\left[\frac{e^{-\imath\psi}\delta(r_{b1})\delta(r_{a2})}{\kappa_{b1}-\kappa_{a2}}+\imath\frac{I_1^{(N)}(r_{b1},e^{\imath\psi}r_{a2})}{2\pi(r_{b1}-e^{\imath\psi}r_{a2})}\chi(\kappa_{b1}-\kappa_{a2})\right]_{1\leq a,b\leq k}\nonumber\\
 \fl&\times&\exp\left(-\imath\Str\kappa r\right)\Phi(r)d[r]\,.\label{6.2.6}
\end{eqnarray}
Again we are able to distinguish the Efetov-Wegner terms from those terms corresponding to supergroup integrals. In the determinant of Eq.~\eref{6.2.6} and in the left upper block of Eq.~\eref{6.2.2}, see Eq.~\eref{6.2.3}, the Dirac-distributions are the contributions from the Efetov-Wegner terms. When we expand the determinants in the Dirac-distributions we obtain the leading terms of $Z_{k}^{(N)}$, $Z_{k-1}^{(N)}$, $\ldots$, $Z_{0}^{(N)}$ which are exactly those found in Refs.~\cite{Guh06,KGG08}. Thus, we have found an expression which can be understood as a generator for all generating functions $Z_{k}^{(N)}$.

The external matrix $H_0$ can also be drawn from another random matrix ensemble as it was done in Refs.~\cite{PanMeh83,Guh96b,Guh96c,Sto02,GuhSto04,Joh07}. However, we do not perform the calculation, here, since it is straightforward to those in an application of Ref.~\cite{KieGuh09a}.

\section{Remarks and conclusions}

We derived the supermatrix Bessel function with all Efetov-Wegner terms for Hermitian supermatrices of arbitrary dimensions. We arrived at an expression from which one can easily deduce what the Efetov-Wegner terms are and which terms result from supergroup integrals. With this result we showed that the completeness and orthogonality relation for the supermatrix Bessel function without Efetov-Wegner terms slightly differs from the formerly assumed one \cite{Guh96b}. It has to be zero on a set of measure zero and, thus, does not matter for smooth integrands but it plays an important role if the integrand has singularities on this set.

We applied the supermatrix Bessel function with Efetov-Wegner terms to arbitrary, rotation invariant Hermitian random matrix ensembles with and without an external field. The already known leading terms \cite{PanMeh83,Guh96b,Guh96c,BreHik98,Sto02,GuhSto04,Guh06,Guh06b,Joh07,KGG08,KieGuh09a} were obtained plus all Efetov-Wegner terms. The correction terms were unknown before in this explicit form and yield new insights in the supersymmetric representation of the generating functions. In particular the Efetov-Wegner terms become important for the matrix Green functions.

We also found an integral identity for the generating functions whose integrand can be easily expanded in the Efetov-Wegner terms. In such an expansion one obtains correlation functions related to $k$-point correlation functions which are of lower order than those corresponding to the originally considered generating function. Thus, it reflects the relation of Mehta's definition \cite{Meh04} for the $k$-point correlation function and the one commonly used in the supersymmetry method \cite{Guh06} which was explained in Ref.~\cite{GMW98}.

We expect that similar results may also be derived for other supermatrices, e.g. diagonalization of complex supermatrices \cite{GuhWet96}. Nevertheless we guess that the knowledge about the supergroup integrals as well as about the ordinary group integrals is crucial. We could only obtain these compact results due to this knowledge.

\section*{Acknowledgements}

I thank Thomas Guhr for fruitful discussions, Taro Nagao for helpful comments about applications as well as Martin Zirnbauer by pointing out the diploma thesis by Ralf Bundschuh. I acknowledge support from the Deutsche Forschungsgemeinschaft within Sonderforschungsbereich Transregio 12 ``Symmetries and Universality in Mesoscopic Systems''.

\appendix

\section{Derivation of Eq.~\eref{3.6}}\label{app1}

We plug the characteristic function~\eref{3.3} in Eq.~\eref{3.1} and diagonalize $\widetilde{H}$. This yields
\begin{eqnarray}
 \fl\displaystyle Z_{k_1/k_2}^{(N)}(\kappa)&=&\frac{V_N}{2^N\pi^{N^2}}\int\limits_{\Herm(N)}\int\limits_{\mathbb{R}^N}\exp\left[-\imath\tr H\widetilde{E}\right]\frac{\prod\limits_{j=1}^{k_2}\det(H-\kappa_{j2}\eins_{N})}{\prod\limits_{j=1}^{k_1}\det(H-\kappa_{j1}\eins_{N})}\prod\limits_{j=1}^Nf(\widetilde{E}_j)\Delta_N^2(\widetilde{E})d[\widetilde{E}]d[H]\,,\nonumber\\
\fl&&\label{a1.1}
\end{eqnarray}
where the constant is
\begin{equation}\label{a1.1b}
 V_N=\frac{1}{N!}\prod\limits_{j=1}^N\frac{\pi^{j-1}}{(j-1)!}\,.
\end{equation}
The diagonalization of $H$ yields the matrix Bessel function \cite{Har58,ItzZub80} according to the unitary group $\U(N)$,
\begin{eqnarray}
 \varphi_{N}(E,\widetilde{E})&=&\int\limits_{\U(N)}\exp\left[-\imath\tr EU\widetilde{E}U^\dagger\right]d\mu(U)\nonumber\\
 &=&\displaystyle\prod\limits_{j=1}^{N}\imath^{j-1}(j-1)!\frac{\det\left[\exp(-\imath E_a\widetilde{E}_b)\right]_{1\leq a,b\leq N}}{\Delta_N(\widetilde{E})\Delta_N(E)}\label{a1.2}
\end{eqnarray}
The measure $d\mu$ is the normalized Haar-measure. Thus, we find
\begin{eqnarray}
 \displaystyle Z_{k_1/k_2}^{(N)}(\kappa)&=&\frac{\imath^{N(N-1)/2}}{(2\pi)^NN!\prod\limits_{j=0}^Nj!}\quad\int\limits_{\mathbb{R}^{2N}}\det\left[\exp(-\imath E_a\widetilde{E}_b)\right]_{1\leq a,b\leq N}\nonumber\\
 &\times&\prod\limits_{a=1}^Nf(\widetilde{E}_a)\frac{\prod\limits_{b=1}^{k_2}(E_a-\kappa_{b2})}{\prod\limits_{b=1}^{k_1}(E_a-\kappa_{b1})}\Delta_N(\widetilde{E})\Delta_N(E)d[\widetilde{E}]d[E]\,.\label{a1.3}
\end{eqnarray}
Please notice that we integrate first over the variables $\widetilde{E}$ and then over $E$. Here, one can easily check that the normalization is $Z_{k_1/k_2}^{(N)}(0)=f^N(0)$. Since determinants are skew-symmetric, we first expand the Vandermonde determinant $\Delta_N(\widetilde{E})$ and then the determinant of the exponential functions. We have
\begin{eqnarray}
 \fl\displaystyle Z_{k_1/k_2}^{(N)}(\kappa)&=&\frac{(-\imath)^{N(N-1)/2}}{(2\pi)^N\prod\limits_{j=1}^N(j-1)!}\quad\int\limits_{\mathbb{R}^{2N}}
 \prod\limits_{a=1}^N\left[f(\widetilde{E}_a)\exp(-\imath E_a\widetilde{E}_a)\widetilde{E}_a^{a-1}\frac{\prod\limits_{b=1}^{k_2}(E_a-\kappa_{b2})}{\prod\limits_{b=1}^{k_1}(E_a-\kappa_{b1})}\right]\nonumber\\
 \fl&\times&\Delta_N(E)d[\widetilde{E}]d[E]\,.\label{a1.4}
\end{eqnarray}
Following the ideas in Ref.~\cite{KieGuh09a}, we extend the integrand by a square root Berezinian and find with help of Eq.~\eref{2.0c} the determinant
\begin{eqnarray}
 \fl\displaystyle Z_{k_1/k_2}^{(N)}(\kappa)&=&\frac{(-1)^{k_2(k_2-1)/2+(k_2+1)k_1}\imath^{N(N-1)/2}}{(2\pi)^N\prod\limits_{j=1}^N(j-1)!}\frac{1}{\sqrt{\Ber_{k_1/k_2}^{(2)}(\kappa)}}\label{a1.5}\\
 \fl&\times&\det\left[\begin{array}{cc} \left\{\displaystyle\frac{1}{\kappa_{a1}-\kappa_{b2}}\right\}\underset{1\leq b\leq k_2}{\underset{1\leq a\leq k_1}{\ }} & \left\{\displaystyle \int\limits_{\mathbb{R}^2}\frac{f(E_2)E_2^{b-1}}{\kappa_{a1}-E_1}\exp[-\imath E_2E_1]d[E]\right\}\underset{1\leq b\leq N}{\underset{1\leq a\leq k_1}{\ }} \\ \left\{\displaystyle\kappa_{b2}^{a-1}\right\}\underset{1\leq b\leq k_2}{\underset{1\leq a\leq d}{\ }} & \left\{\displaystyle \int\limits_{\mathbb{R}^2}f(E_2)E_2^{b-1}E_1^{a-1}\exp[-\imath E_2E_1]d[E]\right\}\underset{1\leq b\leq N}{\underset{1\leq a\leq d}{\ }} \end{array}\right]\,.\nonumber
\end{eqnarray}

We define the sign of the imaginary parts of $\kappa_{j1}$ by
\begin{equation}\label{a1.6}
 -L_j=\frac{\IM\kappa_{j1}}{|\IM\kappa_{j1}|}\,.
\end{equation}
Integrating over $E_1$, Eq.~\eref{a1.5} reads
\begin{eqnarray}
 \fl\displaystyle Z_{k_1/k_2}^{(N)}(\kappa)&=&\frac{(-1)^{k_2(k_2-1)/2+(k_2+1)k_1}\imath^{N(N-1)/2}}{\prod\limits_{j=1}^N(j-1)!}\frac{1}{\sqrt{\Ber_{k_1/k_2}^{(2)}(\kappa)}}\label{a1.7}\\
 \fl&\times&\det\left[\begin{array}{cc} \left\{\displaystyle\frac{1}{\kappa_{a1}-\kappa_{b2}}\right\}\underset{1\leq b\leq k_2}{\underset{1\leq a\leq k_1}{\ }} & \left\{\displaystyle \imath L_a\int\limits_{\mathbb{R}}f(E)E^{b-1}\exp[-\imath\kappa_{a1}E]\Theta(L_aE)dE\right\}\underset{1\leq b\leq N}{\underset{1\leq a\leq k_1}{\ }} \\ \left\{\displaystyle\kappa_{b2}^{a-1}\right\}\underset{1\leq b\leq k_2}{\underset{1\leq a\leq d}{\ }} & \left\{\displaystyle \int\limits_{\mathbb{R}}f(E)E^{b-1}\left(\imath\frac{\partial}{\partial E}\right)^{a-1}\delta(E)dE\right\}\underset{1\leq b\leq N}{\underset{1\leq a\leq d}{\ }} \end{array}\right]\,.\nonumber
\end{eqnarray}
In the lower right block we use the following property of the integral
\begin{equation}\label{a1.8}
 \int\limits_{\mathbb{R}}f(E)E^{b-1}\left(\imath\frac{\partial}{\partial E}\right)^{a-1}\delta(E)dE=0\quad\mathrm{for}\ b>a\,.
\end{equation}
Since $d=N+k_2-k_1\leq N$, cf. Eq.~\eref{3.5}, the last $N-d$ columns in the lower right block in the determinant \eref{a1.7} are zero. The matrix
\begin{equation}\label{a1.9}
 \mathbf{M}=\left[\displaystyle \int\limits_{\mathbb{R}}f(E)E^{b-1}\left(\imath\frac{\partial}{\partial E}\right)^{a-1}\delta(E)dE\right]_{1\leq a,b\leq d}
\end{equation}
is a lower triangular matrix with diagonal elements
\begin{equation}\label{a1.10}
 M_{jj}=\displaystyle \int\limits_{\mathbb{R}}f(E)E^{j-1}\left(\imath\frac{\partial}{\partial E}\right)^{j-1}\delta(E)dE=(-\imath)^{j-1}(j-1)!\,.
\end{equation}
Thus, the determinant of this matrix is
\begin{equation}\label{a1.11}
 \det\mathbf{M}=(-\imath)^{N(N-1)/2}\prod\limits_{j=1}^N(j-1)!\,.
\end{equation}
We pull the Matrix $\mathbf{M}$ out the determinant~\eref{a1.7} and find
\begin{eqnarray}
 \fl\displaystyle &&Z_{k_1/k_2}^{(N)}(\kappa)=\frac{(-1)^{k_2(k_2-1)/2+(k_2+1)k_1}}{\sqrt{\Ber_{k_1/k_2}^{(2)}(\kappa)}}\label{a1.12}\\
 \fl&\times&\det\left[\begin{array}{cc} \left\{\displaystyle K^{(d)}(\kappa_{a1},\kappa_{b2})\right\}\underset{1\leq b\leq k_2}{\underset{1\leq a\leq k_1}{\ }} & \left\{\displaystyle \imath L_a\int\limits_{\mathbb{R}}f(E)E^{b-1}\exp[-\imath\kappa_{a1}E]\Theta(L_aE)dE\right\}\underset{d+1\leq b\leq N}{\underset{1\leq a\leq k_1}{\ }}\end{array}\right]\,,\nonumber
\end{eqnarray}
where
\begin{equation}\label{a1.13}
 \fl\displaystyle K^{(d)}(\kappa_{a1},\kappa_{b2})=\frac{1}{\kappa_{a1}-\kappa_{b2}}-\imath L_a\sum\limits_{m,n=1}^d\int\limits_{\mathbb{R}}f(E)E^{m-1}\exp[-\imath\kappa_{a1}E]\Theta(L_aE)dEM^{-1}_{mn}\kappa_{b2}^{n-1}\,.
\end{equation}
Again we use the fact that the determinant is skew-symmetric which allows also to write
\begin{eqnarray}
 \fl\displaystyle &&Z_{k_1/k_2}^{(N)}(\kappa)=\frac{(-1)^{k_2(k_2-1)/2+(k_2+1)k_1}}{\sqrt{\Ber_{k_1/k_2}^{(2)}(\kappa)}}\label{a1.14}\\
 \fl&\times&\det\left[\begin{array}{cc} \left\{\displaystyle K^{(\widetilde{N})}(\kappa_{a1},\kappa_{b2})\right\}\underset{1\leq b\leq k_2}{\underset{1\leq a\leq k_1}{\ }} & \left\{\displaystyle \imath L_a\int\limits_{\mathbb{R}}f(E)E^{b-1}\exp[-\imath\kappa_{a1}E]\Theta(L_aE)dE\right\}\underset{d+1\leq b\leq N}{\underset{1\leq a\leq k_1}{\ }}\end{array}\right]\,,\nonumber
\end{eqnarray}
for an arbitrary $\widetilde{N}\in\{d,d+1,\ldots,N\}$. For the cases $(k_1/k_2)=(1/1)$ and $(k_1/k_2)=(1/0)$, we identify
\begin{eqnarray}
 Z_{1/1}^{(N)}(\kappa_{a1},\kappa_{b2})&=&\displaystyle(\kappa_{a1}-\kappa_{b2})K^{(N)}(\kappa_{a1},\kappa_{b2})\label{a1.15}\,,\\
 Z_{1/0}^{(N)}(\kappa_{a1})&=&-\displaystyle\imath L_a\int\limits_{\mathbb{R}}f(E)E^{N-1}\exp[-\imath\kappa_{a1}E]\Theta(L_aE)dE\label{a1.16}\,.
\end{eqnarray}
This yields the result~\eref{3.6}.

\section{Derivation of Eq.~\eref{4.13}}\label{app2}

This derivation is similar to the one for the supermatrix Bessel function with Efetov-Wegner term in Sec.~V.A of Ref.~\cite{KKG08}. We consider the integral
\begin{equation}\label{a2.1}
 \fl\frac{Z_{1/1}^{(N)}(\kappa)}{f^N(0)}=\frac{(-1)^{N}2\pi}{(N-1)!}\int\limits_{\Sigma_{1/1}^{(\psi)}}\Phi^{(1/1)}(\hat{\rho})\exp[-\imath\Str\kappa\hat{\rho}]r_1^N\left(e^{-\imath\psi}\frac{\partial}{\partial r_{2}}\right)^{N-1}e^{-\imath\psi}\delta\left(r_{2}\right) d[\rho]\,.
\end{equation}
As in Ref.~\cite{KKG08}, we exchange the integration over the Grassmann variables by a differential operator which yields
\begin{eqnarray}
\fl\frac{Z_{1/1}^{(N)}(\kappa)}{f^N(0)}&=&\frac{(-1)^{N}}{(N-1)!}\int\limits_{\mathbb{R}_+\times\mathbb{R}}r_1^N\left[\left(e^{-\imath\psi}\frac{\partial}{\partial r_{2}}\right)^{N-1}\delta\left(r_{2}\right)\right]\left[\imath\frac{\kappa_1-\kappa_2}{r_1-e^{\imath\psi}r_2}\right.\label{a2.2}\\
\fl&+&\left.\frac{1}{r_1-e^{\imath\psi}r_2}\left(\frac{\partial}{\partial r_1}+e^{-\imath\psi}\frac{\partial}{\partial r_2}\right)-\frac{e^{-\imath\psi}}{r_1}\frac{\partial}{\partial r_2}\right]\left[\frac{f(r_1)}{f\left(e^{\imath\psi}r_2\right)}\exp[-\imath\Str\kappa r]\right]dr_1dr_2\,.\nonumber
\end{eqnarray}
The term
\begin{equation}\label{a2.3}
 \fl Z_1=\frac{\imath(-1)^{N}}{(N-1)!}\int\limits_{\mathbb{R}_+\times\mathbb{R}}\frac{\kappa_1-\kappa_2}{r_1-e^{\imath\psi}r_2}\frac{f(r_1)}{f\left(e^{\imath\psi}r_2\right)}\exp[-\imath\Str\kappa r]r_1^N\left(e^{-\imath\psi}\frac{\partial}{\partial r_{2}}\right)^{N-1}\delta\left(r_{2}\right)dr_1dr_2
\end{equation}
contains the supermatrix Bessel function with respect to $\U(1/1)$ \cite{Guh91,Guh96,KKG08}. The second term
\begin{eqnarray}
 \fl Z_2&=&\frac{(-1)^{N}}{(N-1)!}\int\limits_{\mathbb{R}_+\times\mathbb{R}}r_1^N\left(e^{-\imath\psi}\frac{\partial}{\partial r_{2}}\right)^{N-1}\delta\left(r_{2}\right)\label{a2.4}\\
 \fl&\times&\left[\frac{1}{r_1-e^{\imath\psi}r_2}\left(\frac{\partial}{\partial r_1}+e^{-\imath\psi}\frac{\partial}{\partial r_2}\right)-\frac{e^{-\imath\psi}}{r_1}\frac{\partial}{\partial r_2}\right]\left[\frac{f(r_1)}{f\left(e^{\imath\psi}r_2\right)}\exp[-\imath\Str\kappa r]\right]dr_1dr_2\,.\nonumber
\end{eqnarray}
has to yield the Efetov--Wegner term. By partial integration, we evaluate the Dirac distribution and omit the generalized Wick--rotation. Thus, Eq.~\eref{a2.2} becomes
\begin{eqnarray}
 \fl Z_2&=&\frac{-1}{(N-1)!}\int\limits_{\mathbb{R}_+}\left[\sum\limits_{j=0}^{N-1}\frac{(N-1)!}{j!}r_1^{j}\left(\frac{\partial^{j+1}}{\partial r_1\partial r_2^{j}}+\frac{\partial^{j+1}}{\partial r_2^{j+1}}\right)-r_1^{N-1}\frac{\partial^N}{\partial r_2^N}\right]\label{a2.5}\\
 \fl&\times&\left.\left[\frac{f(r_1)}{f\left(r_2\right)}\exp[-\imath\Str\kappa r]\right]\right|_{r_2=0}dr_1\,.\nonumber
\end{eqnarray}
For all terms up to $j=0$ we perform a partial integration in $r_1$ and find a telescope sum. Hence, we have
\begin{eqnarray}
 Z_2&=&-\int\limits_{\mathbb{R}_+}\left.\frac{\partial}{\partial r_1}\left[\frac{f(r_1)}{f\left(r_2\right)}\exp[-\imath\Str\kappa r]\right]\right|_{r_2=0}dr_1=1\,.\label{a2.6}
\end{eqnarray}
This is indeed the Efetov--Wegner term.

The second equality~\eref{4.13} follows from
\begin{eqnarray}
 \fl&&\displaystyle\int\limits_{\mathbb{R}_+\times\mathbb{R}}\frac{1}{r_1-e^{\imath\psi}r_2}\left(\frac{\partial}{\partial r_1}+e^{\imath\psi}\frac{\partial}{\partial r_2}\right)\left(\frac{f(r_1)}{f\left(e^{\imath\psi}r_2\right)}\exp[-\imath\Str\kappa r]r_1^N\left(e^{-\imath\psi}\frac{\partial}{\partial r_2}\right)^{N-1}\delta(r_2)\right)dr_1dr_2\nonumber\\
 \fl&=&\displaystyle\int\limits_{\mathbb{R}_+\times\mathbb{R}}r_1^N\left[\left(e^{-\imath\psi}\frac{\partial}{\partial r_2}\right)^{N-1}\delta(r_2)\right]\frac{1}{r_1-e^{\imath\psi}r_2}\left(\frac{\partial}{\partial r_1}+e^{\imath\psi}\frac{\partial}{\partial r_2}\right)\left(\frac{f(r_1)}{f\left(e^{\imath\psi}r_2\right)}\exp[-\imath\Str\kappa r]\right)dr_1dr_2\nonumber\\
 \fl&+&\displaystyle\int\limits_{\mathbb{R}_+\times\mathbb{R}}\frac{f(r_1)}{f\left(e^{\imath\psi}r_2\right)}\exp[-\imath\Str\kappa r]\left[\frac{Nr_1^{N-1}}{r_1-e^{\imath\psi}r_2}\left(e^{-\imath\psi}\frac{\partial}{\partial r_2}\right)^{N-1}+\frac{r_1^{N}}{r_1-e^{\imath\psi}r_2}\left(e^{-\imath\psi}\frac{\partial}{\partial r_2}\right)^{N}\right]\delta(r_2)dr_1dr_2\nonumber\\
 \fl&=&\displaystyle\int\limits_{\mathbb{R}_+\times\mathbb{R}}r_1^N\left[\left(e^{-\imath\psi}\frac{\partial}{\partial r_2}\right)^{N-1}\delta(r_2)\right]\frac{1}{r_1-e^{\imath\psi}r_2}\left(\frac{\partial}{\partial r_1}+e^{\imath\psi}\frac{\partial}{\partial r_2}\right)\left(\frac{f(r_1)}{f\left(e^{\imath\psi}r_2\right)}\exp[-\imath\Str\kappa r]\right)dr_1dr_2\nonumber\\
 \fl&+&\displaystyle(-1)^{N-1}\int\limits_{\mathbb{R}_+}\left.\left[\sum\limits_{j=0}^{N-1}\frac{N!}{j!}r_1^{j-1}\frac{\partial^{j}}{\partial r_2^j}-\sum\limits_{j=0}^{N}\frac{N!}{j!}r_1^{j-1}\frac{\partial^{j}}{\partial r_2^j}\right]\left(\frac{f(r_1)}{f\left(e^{\imath\psi}r_2\right)}\exp[-\imath\Str\kappa r]\right)\right|_{r_2=0}dr_1\,.\label{a2.7}
\end{eqnarray}
Both sums cancel each other up to the term $j=N$ which is the term $r_1^{N-1}\partial^N/\partial r_2^N$ in Eq.~\eref{a2.5}.

\section{Derivation of Eq.~\eref{5.6}}\label{app3}

Let the characteristic function and, hence, the superfunction $\Phi^{(k_1/k_2)}$ be factorizable, cf. Eq.~\eref{4.11}. To show the identity~\eref{5.6} we plug Eqs.~\eref{4.3} and \eref{4.13} into the result~\eref{3.6} for $\widetilde{N}=d$. We find
\begin{eqnarray}
 \fl&&\displaystyle\int\limits_{\mathbb{R}_+^{k_1}\times\mathbb{R}^{k_2}}\Ber^{(2)}_{k_1/k_2}(r)\widehat{\varphi}_{k_1/k_2}(r,\kappa)\left[\frac{\prod\limits_{j=1}^{k_1}f(r_{j1})}{\prod\limits_{j=1}^{k_2}f(e^{\imath\psi}r_{j2})}{\det}^dr_1\prod\limits_{j=1}^{k_2}\left(e^{-\imath\psi}\frac{\partial}{\partial r_{j2}}\right)^{d-1}\delta(r_{j2})d[r]\right]\nonumber\\
 \fl&=&\frac{(-1)^{(k_1+k_2)(k_1+k_2-1)/2}(\imath\pi)^{(k_2-k_1)^2/2-(k_1+k_2)/2}}{2^{k_1k_2}\sqrt{\Ber^{(2)}_{k_1/k_2}(\kappa)}}\displaystyle\det\left[\begin{array}{c} \left\{\displaystyle A(\kappa_{b1},\kappa_{a2})\right\}\underset{1\leq b\leq k_1}{\underset{1\leq a\leq k_2}{\ }} \\ \left\{\displaystyle\int\limits_{\mathbb{R}_+}f(r_1)r_1^{a-1}e^{-\imath\kappa_{b1}r_1}dr_1\right\}\underset{1\leq b\leq k_1}{\underset{d+1\leq a\leq N}{\ }} \end{array}\right]\nonumber\\
 \fl&&\label{a3.1}
\end{eqnarray}
with
\begin{eqnarray}\label{a3.2}
 \fl A(\kappa_{b1},\kappa_{a2})&=&\int\limits_{\mathbb{R}_+\times\mathbb{R}}\frac{\exp(-\imath\kappa_{b1}r_1+\imath e^{\imath\psi}\kappa_{b2}r_2)}{(\kappa_{b1}-\kappa_{a2})(r_1-e^{\imath\psi}r_2)}\\
 \fl&\times&\left(\frac{\partial}{\partial r_1}+e^{-\imath\psi}\frac{\partial}{\partial r_2}\right)\left[\frac{f(r_1)}{f(e^{\imath\psi}r_2)}r_1^d\left(e^{-\imath\psi}\frac{\partial}{\partial r_{2}}\right)^{d-1}\delta(r_2)\right]d[r]\,.\nonumber
\end{eqnarray}
The next step is to pull all factors of $f$, the monomials $r_1^d$ and the distribution $\left(e^{-\imath\psi}\partial/\partial r_{2}\right)^{d-1}\delta(r_2)$ out the determinant. Identifing the remaining terms of the integrands for all $f$ we, then, get Eq.~\eref{5.6}.

\section{Derivation of the supermatrix Bessel function with Efetov-Wegner terms}\label{app4}

We use the result of \ref{app3} as an ansatz in Eq.~\eref{5.1}. To prove that this ansatz is indeed the result we are looking for we construct a boundary value problem in a weak sense. We consider the left hand side of Eq.~\eref{5.1} with the supermatrix
\begin{equation}
 \sigma=\left[\begin{array}{cc} \left\{\sigma_{ab1}\right\}\underset{1\leq a,b\leq k_1}{\ } & \left\{\chi_{ba}^*\right\}\underset{1\leq b\leq k_2}{\underset{1\leq a\leq k_1}{\ }} \\ \left\{\chi_{ab}\right\}\underset{1\leq b\leq k_1}{\underset{1\leq a\leq k_2}{\ }} & \left\{\sigma_{ab2}\right\}\underset{1\leq a,b\leq k_2}{\ } \end{array}\right]\label{a4.1}
\end{equation}
with non-zero entries everywhere instead of a diagonal supermatrix $\kappa$. Then the action of the differential operator
\begin{eqnarray}
 \fl\Str\frac{\partial^2}{\partial \sigma^2}&=&\sum\limits_{a=1}^{k_1}\frac{\partial^2}{\partial\sigma_{aa1}^2}+2\sum\limits_{1\leq a<b\leq k_1}\frac{\partial^2}{\partial\sigma_{ab1}\partial\sigma_{ab1}^*}-\sum\limits_{a=1}^{k_2}\frac{\partial^2}{\partial\sigma_{aa2}^2}-2\sum\limits_{1\leq a<b\leq k_2}\frac{\partial^2}{\partial\sigma_{ab2}\partial\sigma_{ab2}^*}\nonumber\\
 &+&2\underset{1\leq b\leq k_1}{\underset{1\leq a\leq k_2}{\sum }}\frac{\partial^2}{\partial\chi_{ab}^*\partial\chi_{ab}}\label{a4.2}
\end{eqnarray}
on the left hand side of Eq.~\eref{5.1} yields
\begin{equation}
 \fl\Str\frac{\partial^2}{\partial \sigma^2}\int\limits_{\widetilde{\Sigma}_{k_1/k_2}^{(\psi)}}F(\rho)\exp[-\imath\Str\sigma\rho]d[\rho]=-\int\limits_{\widetilde{\Sigma}_{k_1/k_2}^{(\psi)}}F(\rho)\Str\rho^2\exp[-\imath\Str\sigma\rho]d[\rho]\label{a4.3}\,.
\end{equation}
Since the integrand is rotation invariant the integral only depends on the eigenvalues of the supermatrix $\sigma$. This leads to a differential equation in the diagonal supermatrix $\kappa$. With the differential operator $\Str\partial^2/\partial \kappa^2$ defined similar to Eq.~\eref{5.18} we have
\begin{eqnarray}
 \fl&&\frac{1}{\sqrt{\Ber^{(2)}_{k_1/k_2}(\kappa)}}\Str\frac{\partial^2}{\partial \kappa^2}\sqrt{\Ber^{(2)}_{k_1/k_2}(\kappa)}\int\limits_{\widetilde{\Sigma}_{k_1/k_2}^{(\psi)}}F(\rho)\exp[-\imath\Str\kappa\rho]d[\rho]\nonumber\\
 \fl&=&-\int\limits_{\widetilde{\Sigma}_{k_1/k_2}^{(\psi)}}F(\rho)\Str\rho^2\exp[-\imath\Str\kappa\rho]d[\rho]\label{a4.5}\,,
\end{eqnarray}
cf. Ref.~\cite{Guh96b}.

The boundaries of $\widetilde{\Sigma}_{k_1/k_2}^{(\psi)}$ are given by $\widetilde{\Sigma}_{k_1-1/k_2-1}^{(\psi)}$ canonically embedded in $\widetilde{\Sigma}_{k_1/k_2}^{(\psi)}$ if one bosonic eigenvalue of a supermatrix in $\widetilde{\Sigma}_{k_1/k_2}^{(\psi)}$ equals to a fermionic one, i.e. there  are two numbers $a\in\{1,\ldots,k_1\}$ and $b\in\{1,\ldots,k_2\}$ with $\kappa_{a1}=\kappa_{b2}$. For these cases we may use the Cauchy-like integral theorems for Hermitian supermatrices \cite{Weg83,ConGro89,KKG08}. Without loss of generality we consider the case $\kappa_{k_11}=\kappa_{k_22}$ and have
\begin{eqnarray}
 \fl\int\limits_{\widetilde{\Sigma}_{k_1/k_2}^{(\psi)}}F(\rho)\exp[-\imath\Str\kappa\rho]d[\rho]=(-1)^{k_1}2^{2-k_1-k_2}\imath\int\limits_{\widetilde{\Sigma}_{k_1-1/k_2-1}^{(\psi)}}F(\rho)\exp\left[-\imath\Str\left.\kappa\right|_{\kappa_{k_11}=\kappa_{k_22}=0}\rho\right]d[\rho]\,.\nonumber\\
 \fl&&\label{a4.6}
\end{eqnarray}
Here we use the same symbol for the restriction of $F$ on $\widetilde{\Sigma}_{k_1-1/k_2-1}^{(\psi)}$.

The boundary condition~\eref{a4.6} for the distribution~\eref{5.8} can be readily checked. For the differential equation~\eref{a4.5} we expand the determinant~\eref{5.8} in $l\leq k_2$ rows and columns in the upper block. Apart from a constant prefactor each term is given by
\begin{eqnarray}
 g(\kappa,r)&=&\frac{\sqrt{\Ber^{(2)}_{l/l}(\kappa_{\tilde{\omega}})}\sqrt{\Ber^{(2)}_{k_1-l/k_2-l}(r_{\omega})}}{\sqrt{\Ber^{(2)}_{k_1/k_2}(\kappa)}\Ber^{(2)}_{k_1/k_2}(r)}\prod\limits_{j=1}^le^{-\imath\psi}\delta(r_{\omega_1(j)1})\delta(r_{\omega_2(j)2})\nonumber\\
 &\times&\prod\limits_{a=l+1}^{k_1}\exp(-\imath\kappa_{\tilde{\omega}_1(a)1}r_{\omega_1(a)1})\prod\limits_{a=l+1}^{k_2}\exp(\imath e^{\imath\psi}\kappa_{\tilde{\omega}_2(b)2}r_{\omega_2(b)2})\label{a4.7}\,,
\end{eqnarray}
where $\kappa_{\tilde{\omega}}=\diag(\kappa_{\tilde{\omega}_1(1)1},\ldots,\kappa_{\tilde{\omega}_1(l)1},\kappa_{\tilde{\omega}_2(1)2},\ldots,\kappa_{\tilde{\omega}_2(l)2})$ and $r_{\omega}=\diag(r_{\omega_1(l+1)1},\ldots,r_{\omega_1(k_1)1},$ $r_{\omega_2(l+1)2},\ldots,r_{\omega_2(k_2)1})$ with the permutations $\omega_1,\tilde{\omega}_1\in\mathfrak{S}(k_1)$ and $\omega_2,\tilde{\omega}_2\in\mathfrak{S}(k_2)$. The action of the distribution $g(\kappa,r)$ on $\Str r^2$ is
\begin{eqnarray}
 \fl g(\kappa,r)\Str r^2&=&\frac{\sqrt{\Ber^{(2)}_{l/l}(\kappa_{\tilde{\omega}})}\sqrt{\Ber^{(2)}_{k_1-l/k_2-l}(r_{\omega})}}{\sqrt{\Ber^{(2)}_{k_1/k_2}(\kappa)}\Ber^{(2)}_{k_1/k_2}(r)}\prod\limits_{j=1}^le^{-\imath\psi}\delta(r_{\omega_1(j)1})\delta(r_{\omega_2(j)2})\nonumber\\
 \fl&\times&\prod\limits_{a=l+1}^{k_1}\exp(-\imath\kappa_{\tilde{\omega}_1(a)1}r_{\omega_1(a)1})\prod\limits_{a=l+1}^{k_2}\exp(\imath e^{\imath\psi}\kappa_{\tilde{\omega}_2(b)2}r_{\omega_2(b)2})\Str r^2_{\omega}\label{a4.8}
\end{eqnarray}
because all other terms are zero due to the Dirac-distributions. The differential operator in Eq.~\eref{a4.5} acts on $g(\kappa,r)$ as
\begin{eqnarray}
 \fl&&\frac{1}{\sqrt{\Ber^{(2)}_{k_1/k_2}(\kappa)}}\Str\frac{\partial^2}{\partial \kappa^2}\sqrt{\Ber^{(2)}_{k_1/k_2}(\kappa)}g(\kappa,r)\nonumber\\
 \fl&=&\frac{1}{\sqrt{\Ber^{(2)}_{k_1/k_2}(\kappa)}}\Str\frac{\partial^2}{\partial \kappa^2}\frac{\sqrt{\Ber^{(2)}_{l/l}(\kappa_{\tilde{\omega}})}\sqrt{\Ber^{(2)}_{k_1-l/k_2-l}(r_{\omega})}}{\Ber^{(2)}_{k_1/k_2}(r)}\prod\limits_{j=1}^le^{-\imath\psi}\delta(r_{\omega_1(j)1})\delta(r_{\omega_2(j)2})\nonumber\\
 \fl&\times&\prod\limits_{a=l+1}^{k_1}\exp(-\imath\kappa_{\tilde{\omega}_1(a)1}r_{\omega_1(a)1})\prod\limits_{a=l+1}^{k_2}\exp(\imath e^{\imath\psi}\kappa_{\tilde{\omega}_2(b)2}r_{\omega_2(b)2})\,.\label{a4.9}
\end{eqnarray}
We split the differential operator $\Str\partial^2/\partial\kappa^2$ into a part acting on $\kappa_{\tilde{\omega}}$ and a part for the remaining variables in $\kappa$. The term for the latter variables acts on the exponential functions in Eq.~\eref{a4.9} and contributes the term $-\Str r_{\omega}^2$. For the term according to $\kappa_{\tilde{\omega}}$ we use the identity
\begin{equation}
 \Str\frac{\partial^2}{\partial \kappa_{\tilde{\omega}}^2}\sqrt{\Ber^{(2)}_{l/l}(\kappa_{\tilde{\omega}})}=0\,.\label{a4.10}
\end{equation}
Thus, the differential equation is also fulfilled by $\widehat{\varphi}_{k_1/k_2}$.

\section{Double Fourier-transform}\label{app5}

We consider the integral
\begin{eqnarray}
 \fl I&=&\int\limits_{\mathbb{R}^{k_1+k_2}}\det\left[\begin{array}{c} \left\{\displaystyle \frac{-2\pi e^{-\imath\psi}\delta(r_{b1})\delta(r_{a2})}{\kappa_{b1}-e^{-\imath\psi}\kappa_{a2}}+\frac{\exp\left(-\imath\kappa_{b1}r_{b1}+\imath \kappa_{a2}r_{a2}\right)}{r_{b1}-e^{\imath\psi}r_{a2}}\chi(\kappa_{b1}-e^{-\imath\psi}\kappa_{a2})\right\}\underset{1\leq b\leq k_1}{\underset{1\leq a\leq k_2}{\ }} \\ \left\{\displaystyle r_{b1}^{a-1}\exp\left(-\imath\kappa_{b1}r_{b1}\right)\right\}\underset{1\leq b\leq k_1}{\underset{1\leq a\leq k_1-k_2}{\ }} \end{array}\right]\nonumber\\
 \fl&\times&\det\left[\begin{array}{c} \left\{\displaystyle \frac{2\pi e^{\imath\psi}\delta(s_{b1})\delta(s_{a2})}{r_{b1}-e^{\imath\psi}r_{a2}}-\frac{\exp\left(\imath r_{b1}s_{b1}-\imath r_{a2}s_{a2}\right)}{s_{b1}-e^{-\imath\psi}s_{a2}}\chi(r_{b1}-e^{\imath\psi}r_{a2})\right\}\underset{1\leq b\leq k_1}{\underset{1\leq a\leq k_2}{\ }} \\ \left\{\displaystyle s_{b1}^{a-1}\exp\left(\imath r_{b1}s_{b1}\right)\right\}\underset{1\leq b\leq k_1}{\underset{1\leq a\leq k_1-k_2}{\ }} \end{array}\right]\nonumber\\
 \fl&\times&\frac{d[r]}{\sqrt{\Ber^{(2)}_{k_1/k_2}(r)}\sqrt{\Ber^{(2)}_{k_1/k_2}(\widetilde{\kappa)}}\Ber^{(2)}_{k_1/k_2}(s)}\,.\label{a5.1}
\end{eqnarray}
We omit the two sums over the permutation groups, see Eq.~\eref{5.8}. They do not contribute any additional new information of the calculation and the missing terms can be regained by permuting the indices of the eigenvalues in $s$ or $\widetilde{\kappa}$.

The expansion in the first determinant yields
\begin{eqnarray}
 \fl I&=&\sum\limits_{l=0}^{k_2}\underset{\omega_2\in\mathfrak{S}(k_2)}{\underset{\omega_1\in\mathfrak{S}(k_1)}{\sum}}\frac{\sign\omega_1\omega_2}{(l!)^2(k_1-l)!(k_2-l)!}\int\limits_{\mathbb{R}^{k_1+k_2}}\det\left[\frac{-2\pi e^{-\imath\psi}\delta(r_{\omega_1(b)1})\delta(r_{\omega_2(a)2})}{\kappa_{\omega_1(b)1}-e^{-\imath\psi}\kappa_{\omega_2(a)2}}\right]_{1\leq a,b\leq l}\nonumber\\
 \fl&\times&\det\left[\begin{array}{c} \left\{\displaystyle \frac{\exp\left(-\imath\kappa_{\omega_1(b)1}r_{\omega_1(b)1}+\imath \kappa_{\omega_2(a)2}r_{\omega_2(a)2}\right)}{r_{\omega_1(b)1}-e^{\imath\psi}r_{\omega_2(a)2}}\chi(\kappa_{\omega_1(b)1}-e^{-\imath\psi}\kappa_{\omega_2(a)2})\right\}\underset{l+1\leq b\leq k_1}{\underset{l+1\leq a\leq k_2}{\ }} \\ \left\{\displaystyle r_{\omega_1(b)1}^{a-1}\exp\left(-\imath\kappa_{\omega_1(b)1}r_{\omega_1(b)1}\right)\right\}\underset{l+1\leq b\leq k_1}{\underset{1\leq a\leq k_1-k_2}{\ }} \end{array}\right]\nonumber\\
 \fl&\times&\det\left[\begin{array}{c} \left\{\displaystyle \frac{2\pi e^{\imath\psi}\delta(s_{b1})\delta(s_{a2})}{r_{b1}-e^{\imath\psi}r_{a2}}-\frac{\exp\left(\imath r_{b1}s_{b1}-\imath r_{a2}s_{a2}\right)}{s_{b1}-e^{-\imath\psi}s_{a2}}\chi(r_{b1}-e^{\imath\psi}r_{a2})\right\}\underset{1\leq b\leq k_1}{\underset{1\leq a\leq k_2}{\ }} \\ \left\{\displaystyle s_{b1}^{a-1}\exp\left(\imath r_{b1}s_{b1}\right)\right\}\underset{1\leq b\leq k_1}{\underset{1\leq a\leq k_1-k_2}{\ }} \end{array}\right]\nonumber\\
 \fl&\times&\frac{d[r]}{\sqrt{\Ber^{(2)}_{k_1/k_2}(r)}\sqrt{\Ber^{(2)}_{k_1/k_2}(\widetilde{\kappa)}}\Ber^{(2)}_{k_1/k_2}(s)}\nonumber\\
 \fl&=&(-1)^{k_1(k_1-1)/2}\sum\limits_{l=0}^{k_2}\underset{\omega_2\in\mathfrak{S}(k_2)}{\underset{\omega_1\in\mathfrak{S}(k_1)}{\sum}}\frac{\sign\omega_1\omega_2}{(l!)^2(k_1-l)!(k_2-l)!}\det\left[\frac{-(2\pi)^2e^{\imath\psi}\delta(s_{\omega_1(b)1})\delta(s_{\omega_2(a)2})}{\kappa_{\omega_1(b)1}-e^{-\imath\psi}\kappa_{\omega_2(a)2}}\right]_{1\leq a,b\leq l}\nonumber\\
 \fl&\times&\prod\limits_{a=l+1}^{k_1}\prod\limits_{b=l+1}^{k_2}\chi(\kappa_{\omega_1(b)1}-e^{-\imath\psi}\kappa_{\omega_2(a)2})\int\limits_{\mathbb{R}^{k_1+k_2-2l}}\prod\limits_{a=l+1}^{k_1}\exp\left(-\imath\kappa_{\omega_1(a)1}r_{a1}\right)\prod\limits_{b=l+1}^{k_2}\exp\left(\imath \kappa_{\omega_2(b)2}r_{b2}\right)\nonumber\\
 \fl&\times&\det\left[\begin{array}{c} \left\{\displaystyle \frac{2\pi e^{\imath\psi}\delta(s_{\omega_1(b)1})\delta(s_{\omega_2(a)2})}{r_{b1}-e^{\imath\psi}r_{a2}}-\frac{\exp\left(\imath r_{b1}s_{\omega_1(b)1}-\imath r_{a2}s_{\omega_2(a)2}\right)}{s_{\omega_1(b)1}-e^{-\imath\psi}s_{\omega_2(a)2}}\chi(r_{b1}-e^{\imath\psi}r_{a2})\right\}\underset{l+1\leq b\leq k_1}{\underset{l+1\leq a\leq k_2}{\ }} \\ \left\{\displaystyle s_{\omega_1(b)1}^{a-1}\exp\left(\imath r_{b1}s_{\omega_1(b)1}\right)\right\}\underset{l+1\leq b\leq k_1}{\underset{1\leq a\leq k_1-k_2}{\ }} \end{array}\right]\nonumber\\
 \fl&\times&\frac{d[r]}{\sqrt{\Ber^{(2)}_{k_1/k_2}(\widetilde{\kappa)}}\Ber^{(2)}_{k_1/k_2}(s)}\label{a5.2}
\end{eqnarray}
where the function ``$\sign$'' yields $1$ for an even permutation and $-1$ for an odd one. The permutations in the indices of the $r$ are absorbed in the integration. We remark that the remaining integral goes over $k_1+k_2-2l$ variables because we have already used the Dirac-distributions.

With help of the formula
\begin{eqnarray}
 \fl&&\int_{\mathbb{R}^2}\frac{2\pi e^{\imath\psi}\delta(s_{\omega_1(b)1})\delta(s_{\omega_2(a)2})}{r_{b1}-e^{\imath\psi}r_{a2}}\exp\left(-\imath\kappa_{\omega_1(b)1}r_{b1}+\imath \kappa_{\omega_2(a)2}r_{a2}\right)d[r]\nonumber\\
 \fl&=&\int_{\mathbb{R}^2}\frac{2\pi\imath e^{\imath\psi}\delta(s_{\omega_1(b)1})\delta(s_{\omega_2(a)2})}{(\kappa_{\omega_1(b)1}-e^{-\imath\psi}\kappa_{\omega_2(a)2})(r_{b1}-e^{\imath\psi}r_{a2})}\nonumber\\
 \fl&\times&\left(\frac{\partial}{\partial r_{b1}}+e^{-\imath\psi}\frac{\partial}{\partial r_{a2}}\right)\exp\left(-\imath\kappa_{\omega_1(b)1}r_{b1}+\imath \kappa_{\omega_2(a)2}r_{a2}\right)d[r]\nonumber\\
 \fl&=&\frac{(2\pi)^2 e^{\imath\psi}\delta(s_{\omega_1(b)1})\delta(s_{\omega_2(a)2})}{\kappa_{\omega_1(b)1}-e^{-\imath\psi}\kappa_{\omega_2(a)2}}\label{a5.3}
\end{eqnarray}
we integrate and sum the expression~\eref{a5.2} up. This yields
\begin{eqnarray}\label{a5.4}
 \fl&&\int\limits_{\mathbb{R}^{k_1+k_2}}\widehat{\varphi}_{k_1/k_2}(\imath s,r)\widehat{\varphi}_{k_1/k_2}(-\imath r,\widetilde{\kappa})\Ber^{(2)}_{k_1/k_2}(r)d[r]\\
 \fl&=&\frac{(-1)^{k_1(k_1-1)/2}\pi^{(k_2-k_1)^2}}{2^{2k_1k_2-k_1-k_2}k_1!k_2!\sqrt{\Ber_{k_1/k_2}^{(2)}(\widetilde{\kappa})}\Ber_{k_1/k_2}^{(2)}(s)}\underset{\omega_2\in\mathfrak{S}(k_2)}{\underset{\omega_1\in\mathfrak{S}(k_1)}{\sum}}\det\left[\begin{array}{c} \underset{\ }{\displaystyle B_{ab}} \\  \displaystyle \overset{\ }{s_{\omega_1(b)1}^{a-1}\delta(\kappa_{b1}-s_{\omega_1(b)1})} \end{array}\right]\nonumber
\end{eqnarray}
with
\begin{eqnarray}
 B_{ab}&=&\frac{e^{\imath\psi}\delta(s_{\omega_1(b)1})\delta(s_{\omega_2(a)2})}{\kappa_{b1}-e^{-\imath\psi}\kappa_{a2}}[1-\chi(\kappa_{b1}-e^{-\imath\psi}\kappa_{a2})]\nonumber\\
   &+&e^{\imath\psi}\frac{\delta(\kappa_{b1}-s_{\omega_1(b)1})\delta(\kappa_{a2}-s_{\omega_2(a)2})}{s_{\omega_1(b)1}-e^{-\imath\psi}s_{\omega_2(a)2}}\chi(\kappa_{b1}-e^{-\imath\psi}\kappa_{a2})\,.\label{a5.5}
\end{eqnarray}
This is the result~\eref{5.15}. The index $a$ goes from $1$ to $k_2$ in the upper block and from $1$ to $k_1-k_2$ in the lower block whereas $b$ takes the values from $1$ to $k_1$ in both blocks.

\section{Calculations for subsection~\ref{sec6.1}}\label{app6}

We diagonalize the supermatrices $\sigma$ and $\rho$ in Eq.~\eref{6.1.4} and have for the generating function 
\begin{eqnarray}
 \fl Z_{k}^{(N)}(\kappa)&=&\frac{1}{(2\pi\imath)^{2k}}\int\limits_{\mathbb{R}^{4k}}\frac{d[r]d[s]}{\sqrt{\Ber^{(2)}_{k/k}(r)}\sqrt{\Ber^{(2)}_{k/k}(\kappa)}}\mathcal{F}\Phi(s)I_k^{(N)}(r)\label{a6.1}\\
 \fl&\times&\det\left[\displaystyle \frac{-2\pi e^{-\imath\psi}\delta(r_{b1})\delta(r_{a2})}{\kappa_{b1}-\kappa_{a2}}+\frac{\exp\left(-\imath\kappa_{b1}r_{b1}+\imath e^{\imath\psi} \kappa_{a2}r_{a2}\right)}{r_{b1}-e^{\imath\psi}r_{a2}}\chi(\kappa_{b1}-\kappa_{a2})\right]_{1\leq a,b\leq k}\nonumber\\
 \fl&\times&\det\left[\displaystyle \frac{2\pi e^{\imath\psi}\delta(s_{b1})\delta(s_{a2})}{r_{b1}-e^{\imath\psi}r_{a2}}-\frac{\exp\left(\imath r_{b1}s_{b1}-\imath r_{a2}s_{a2}\right)}{s_{b1}-e^{-\imath\psi}s_{a2}}\chi(r_{b1}-e^{\imath\psi}r_{a2})\right]_{1\leq a,b\leq k}\,.\nonumber
\end{eqnarray}
This expression is not well defined because the supersymmetric Ingham-Siegel is at zero not well defined. We recall that the supersymmetric Ingham-Siegel integral factorizes in each eigenvalue of the supermatrix $r$, cf. Eq.~\eref{6.1.5}. To understand Eq.~\eref{a6.1} we have to know what $I_1^{(N)}(0)$ is.  Since the supersymmetric Ingham-Siegel integral is a distribution we consider an arbitrary rotation invariant, sufficiently integrable superfunction $f$ on the set of $(1+1)\times(1+1)$ Hermitian supermatrices. Then we have
\begin{eqnarray}
 \fl\int\limits_{\widetilde{\Sigma}_{1/1}^{(\psi)}}f(\rho)I_1^{(N)}(\rho)d[\rho]&=&\int\limits_{\widetilde{\Sigma}_{1/1}^{(-\psi)}}\left(\int\limits_{\widetilde{\Sigma}_{1/1}^{(\psi)}}f(\rho)\exp(-\imath\Str\rho\sigma+\varepsilon\Str\rho)d[\rho]\right){\Sdet}^{-N}(\sigma+\imath\varepsilon\eins_{1+1})d[\sigma]\nonumber\\
 \fl&=&\imath\int\limits_{\widetilde{\Sigma}_{1/1}^{(\psi)}}f(\rho)\exp(\varepsilon\Str\rho)d[\rho]\nonumber\\
 \fl&=&f(0)\nonumber\\
 \fl&\overset{!}{=}&-\imath f(0)I_1^{(N)}(0)\label{a6.2}
\end{eqnarray}
with help of the Cauchy-like integral theorem for $(1+1)\times(1+1)$ Hermitian supermatrices, see Ref.~\cite{Weg83,KKG08}. Please notice that the constant resulting from the Cauchy-like integral theorem converts to the complex conjugate when the generalized Wick-rotation is complex conjugated. The last equality in Eq.~\eref{a6.2} is the Cauchy-like integral theorem formally applied to the left hand side of Eq.~\eref{a6.2}. Hence we conclude that $I_1^{(N)}(0)=\imath$ in a distributional sense. Using this result we find
\begin{eqnarray}
\fl &&Z_{k}^{(N)}(\kappa)=\frac{(-1)^{k(k+1)/2}}{2^{k(k+1)}\pi^{2k}}\sum\limits_{l=0}^{k}\sum\limits_{\omega_1,\omega_2\in\mathfrak{S}(k)}\frac{\sign\omega_1\omega_2}{\left[l!(k-l)!\right]^2}\int\limits_{\mathbb{R}^{4k}}\mathcal{F}\Phi(s)\label{a6.3}\\
 \fl&\times&\det\left[\displaystyle \frac{\exp\left(-\imath\kappa_{\omega_1(b)1}r_{\omega_1(b)1}+\imath e^{\imath\psi} \kappa_{\omega_2(a)2}r_{\omega_2(a)2}\right)}{r_{\omega_1(b)1}-e^{\imath\psi}r_{\omega_2(a)2}}I_1^{(N)}(r_{\omega_1(b)1},e^{\imath\psi}r_{\omega_2(a)2})\chi(\kappa_{\omega_1(b)1}-\kappa_{\omega_2(a)2})\right]_{l+1\leq a, b\leq k}\nonumber\\
 \fl&\times&\det\left[\displaystyle \frac{2\pi e^{\imath\psi}\delta(s_{b1})\delta(s_{a2})}{r_{b1}-e^{\imath\psi}r_{a2}}-\frac{\exp\left(\imath r_{b1}s_{b1}-\imath r_{a2}s_{a2}\right)}{s_{b1}-e^{-\imath\psi}s_{a2}}\chi(r_{b1}-e^{\imath\psi}r_{a2})\right]_{1\leq a,b\leq k}\nonumber\\
 \fl&\times&\det\left[\frac{-2\pi\imath e^{-\imath\psi}\delta(r_{\omega_1(b)1})\delta(r_{\omega_2(a)2})}{\kappa_{\omega_1(b)1}-\kappa_{\omega_2(a)2}}\right]_{1\leq a,b\leq l}\frac{d[r]d[s]}{\sqrt{\Ber^{(2)}_{k_1/k_2}(r)}\sqrt{\Ber^{(2)}_{k/k}(\kappa)}}\nonumber\\
 \fl&=&\frac{(-1)^k}{2^{k(k+1)}\pi^{2k}}\sum\limits_{l=0}^{k}\sum\limits_{\omega_1,\omega_2\in\mathfrak{S}(k)}\frac{\sign\omega_1\omega_2}{\left[l!(k-l)!\right]^2}\prod\limits_{a,b=l+1}^{k}\chi(\kappa_{\omega_1(b)1}-\kappa_{\omega_2(a)2})\int\limits_{\mathbb{R}^{4k-2l}}\mathcal{F}\Phi(s)\nonumber\\
 \fl&\times&\prod\limits_{a,b=l+1}^{k}\exp\left(-\imath\kappa_{\omega_1(b)1}r_{b1}+\imath e^{\imath\psi} \kappa_{\omega_2(a)2}r_{a2}\right)I_1^{(N)}(r_{b1},e^{\imath\psi}r_{a2})\nonumber\\
 \fl&\times&\det\left[\displaystyle \frac{2\pi e^{\imath\psi}\delta(s_{\omega_1(b)1})\delta(s_{\omega_2(a)2})}{r_{b1}-e^{\imath\psi}r_{a2}}-\frac{\exp\left(\imath r_{b1}s_{\omega_1(b)1}-\imath r_{a2}s_{\omega_2(a)2}\right)}{s_{\omega_1(b)1}-e^{-\imath\psi}s_{\omega_2(a)2}}\chi(r_{b1}-e^{\imath\psi}r_{a2})\right]_{l+1\leq a,b\leq k}\nonumber\\
 \fl&\times&\det\left[\frac{-(2\pi)^2\imath e^{\imath\psi}\delta(s_{\omega_1(b)1})\delta(s_{\omega_2(a)2})}{\kappa_{\omega_1(b)1}-\kappa_{\omega_2(a)2}}\right]_{1\leq a,b\leq l}\frac{d[r]d[s]}{\sqrt{\Ber^{(2)}_{k/k}(\kappa)}}\nonumber\\
 \fl&=&\frac{1}{2^{k(k-1)}\imath^k}\sum\limits_{l=0}^{k}\sum\limits_{\omega_1,\omega_2\in\mathfrak{S}(k)}\frac{\sign\omega_1\omega_2}{\left[l!(k-l)!\right]^2}\prod\limits_{a,b=l+1}^{k}\chi(\kappa_{\omega_1(b)1}-\kappa_{\omega_2(a)2})\nonumber\\
 \fl&\times&\int\limits_{\mathbb{R}^{2k}}\frac{d[r]d[s]}{\sqrt{\Ber^{(2)}_{k/k}(\kappa)}}\mathcal{F}\Phi(s)\det\left[-\frac{e^{\imath\psi}\delta(s_{\omega_1(b)1})\delta(s_{\omega_2(a)2})}{\kappa_{\omega_1(b)1}-\kappa_{\omega_2(a)2}}\right]_{1\leq a,b\leq l}\nonumber\\
 \fl&\times&\det\left[\displaystyle \frac{e^{\imath\psi}\delta(s_{\omega_1(b)1})\delta(s_{\omega_2(a)2})}{\kappa_{\omega_1(b)1}-\kappa_{\omega_2(a)2}}\left[1-\left(\frac{\kappa_{\omega_2(a)2}}{\kappa_{\omega_1(b)1}}\right)^N\right]\right.\nonumber\\
 \fl&&\left.-\frac{N\left(e^{-\imath\psi}s_{\omega_2(a)2}-\kappa_{\omega_2(a)2}\right)^{N-1}}{2\pi\imath\left(e^{-\imath\psi}s_{\omega_1(b)1}-\kappa_{\omega_1(b)1}\right)^{N+1}(s_{\omega_1(b)1}-e^{-\imath\psi}s_{\omega_2(a)2})}\right]_{l+1\leq a,b\leq k}\nonumber
\end{eqnarray}
We perform the sum and use the identity
\begin{equation}\label{a6.4}
 1-\left[1-\left(\frac{\kappa_{\omega_2(a)2}}{\kappa_{\omega_1(b)1}}\right)^N\right]\chi(\kappa_{\omega_1(b)1}-\kappa_{\omega_2(a)2})=\left(\frac{\kappa_{\omega_2(a)2}}{\kappa_{\omega_1(b)1}}\right)^N\,.
\end{equation}
Then we have the result~\eref{6.1.6}.

\section{Calculations for subsection~\ref{sec6.2}}\label{app7}

In the first step we derive the Fourier-transform of the superdeterminant in Eq.~\eref{6.2.1}. Let the entries of the diagonal $(k+k)\times(k+k)$ supermatrix $r$ and the entries of diagonal $N\times N$ matrix $E_0$ be the eigenvalues of the supermatrix $\rho$ and the Hermitian matrix $H_0$, i.e. $\rho=UrU^\dagger$ with $U\in\U(k/k)$ and $H_0=VE_0V^\dagger$ with $V\in\U(N)$. Then the Fourier-transform is
\begin{eqnarray}
 \fl J&=&\int\limits_{\widetilde{\Sigma}_{k/k}^{(-\psi)}}\exp(-\imath\Str r\sigma+\varepsilon\Str\rho){\Sdet}^{-1}(\sigma\otimes\eins_N+ \alpha\eins_{k+k}\otimes H_0+\imath\varepsilon\eins_{N(k+k)})d[\sigma]\nonumber\\
 \fl&=&\frac{\imath^k}{2^{k^2}\pi^k}\int\limits_{\mathbb{R}^{2k}}\det\left[\frac{2\pi e^{\imath\psi}\delta(s_{b1})\delta(s_{a2})}{r_{b1}-e^{\imath\psi}r_{a2}}+\frac{\exp\left(-\imath s_{b1}r_{b1}+\imath s_{a2}r_{a2}\right)}{s_{b1}-e^{-\imath\psi}s_{a2}}\chi(r_{b1}-e^{\imath\psi}r_{a2})\right]_{1\leq a,b\leq k}\nonumber\\
 \fl&\times&{\Sdet}^{-1}(s\otimes\eins_N+ \alpha\eins_{k+k}\otimes H_0+\imath\varepsilon\eins_{N(k+k)})\frac{d[s]}{\sqrt{\Ber^{(2)}_{k/k}(r)}}\,.\label{a7.1}
\end{eqnarray}
With help of identity~\eref{2.0b} we find
\begin{eqnarray}
 \fl J&=&\frac{\imath^k}{2^{k^2}\pi^k}\frac{\exp(\varepsilon\Str r)}{\sqrt{\Ber^{(2)}_{k/k}(r)}}\sum\limits_{l=0}^{k}\sum\limits_{\omega_1,\omega_2\in\mathfrak{S}(k)}\frac{\sign\omega_1\omega_2}{\left[l!(k-l)!\right]^2}\prod\limits_{a,b=l+1}^k\chi(r_{\omega_1(b)1}-e^{\imath\psi}r_{\omega_2(a)2})\nonumber\\
 \fl&\times&\int\limits_{\mathbb{R}^{2k}}{\Sdet}^{-1}(s\otimes\eins_N+ \alpha\eins_{k+k}\otimes E_0+\imath\varepsilon\eins_{N(k+k)})\det\left[\frac{2\pi e^{\imath\psi}\delta(s_{b1})\delta(s_{a2})}{r_{\omega_1(b)1}-e^{\imath\psi}r_{\omega_2(a)2}}\right]_{1\leq a,b\leq l}\nonumber\\
 \fl&\times&\det\left[\frac{\exp\left(-\imath s_{b1}r_{\omega_1(b)1}+\imath s_{a2}r_{\omega_2(a)2}\right)}{s_{b1}-e^{-\imath\psi}s_{a2}}\right]_{l+1\leq a,b\leq k}d[s]\nonumber\\
 \fl&=&\frac{\imath^k}{2^{k^2}\pi^k}\frac{\exp(\varepsilon\Str r)}{\Delta_N(\alpha E_0)\sqrt{\Ber^{(2)}_{k/k}(r)}}\sum\limits_{l=0}^{k}\sum\limits_{\omega_1,\omega_2\in\mathfrak{S}(k)}\frac{\sign\omega_1\omega_2}{\left[l!(k-l)!\right]^2}\prod\limits_{a,b=l+1}^k\chi(r_{\omega_1(b)1}-e^{\imath\psi}r_{\omega_2(a)2})\nonumber\\
 \fl&\times&\int\limits_{\mathbb{R}^{2k}}\det\left[\frac{2\pi e^{\imath\psi}\delta(s_{b1})\delta(s_{a2})}{r_{\omega_1(b)1}-e^{\imath\psi}r_{\omega_2(a)2}}\right]_{1\leq a,b\leq l}\prod\limits_{j=l+1}^{k}\exp\left(\imath s_{j2}r_{\omega_2(j)2}\right)\label{a7.2}\\
 \fl&\times&\det\left[\begin{array}{c|c}\displaystyle\underset{\ }{\frac{\exp\left(-\imath s_{b1}r_{\omega_1(b)1}\right)}{s_{b1}-e^{-\imath\psi}s_{a2}}\left(\frac{e^{-\imath\psi}s_{a2}+\imath\varepsilon}{s_{b1}+\imath\varepsilon}\right)^N} & \displaystyle\left(e^{-\imath\psi}s_{a2}+\imath\varepsilon\right)^{b-1} \\ \hline \displaystyle\overset{\ }{\frac{\exp\left(-\imath s_{b1}r_{\omega_1(b)1}\right)}{s_{b1}+\imath\varepsilon+\alpha E_{a}^{(0)}}\left(\frac{-\alpha E_{a}^{(0)}}{s_{b1}+\imath\varepsilon}\right)^N} & \displaystyle(-\alpha E_{a}^{(0)})^{b-1} \end{array}\right]d[s]\,.\nonumber
\end{eqnarray}
In the left upper block both indices $a$ and $b$ run from $l+1$ to $k$ whereas in the right lower block the range is from $1$ to $N$. In the right upper block $a$ goes from $l+1$ to $k$ and $b$ goes from $1$ to $N$ whereas it is vice versa in the left lower block. We sum all terms in Eq.~\eref{a7.2} up and pull the integrations into the determinant. Then we have
\begin{eqnarray}
 \fl J&=&\frac{\imath^k}{2^{k^2}\pi^k}\frac{1}{\Delta_N(\alpha E_0)\sqrt{\Ber^{(2)}_{k/k}(r)}}\det\left[\begin{array}{cc}\displaystyle\left\{A_1(r_{b1},e^{\imath\psi}r_{a2})\right\}_{1\leq a,b\leq k} & \displaystyle\left\{A_{b2}(e^{\imath\psi}r_{a2})\right\}\underset{1\leq b\leq N}{\underset{1\leq a\leq k}{\ }} \\  \displaystyle\left\{A_3(r_{b1},\alpha E_a^{(0)})\right\}\underset{1\leq b\leq k}{\underset{1\leq a\leq N}{\ }} & \displaystyle\left\{(-\alpha E_{a}^{(0)})^{b-1}\right\}_{1\leq a,b\leq N} \end{array}\right]\,,\nonumber\\
 \fl&&\label{a7.3}
\end{eqnarray}
where
\begin{eqnarray}
 \fl A_1(r_{b1},e^{\imath\psi}r_{a2})&=&\exp\left[\varepsilon(r_{b1}-e^{\imath\psi}r_{a2})\right]\Biggl(\frac{2\pi}{r_{b1}-e^{\imath\psi}r_{a2}}\nonumber\\
 \fl&+&\int\limits_{\mathbb{R}^2}\frac{\exp\left(-\imath s_{1}r_{b1}+\imath s_{2}r_{a2}\right)}{s_{1}-e^{-\imath\psi}s_{2}}\left(\frac{e^{-\imath\psi}s_{2}+\imath\varepsilon}{s_{1}+\imath\varepsilon}\right)^N\chi(r_{b1}-e^{\imath\psi}r_{a2})d[s]\Biggl)\nonumber\\
 \fl&=&-2\pi\imath\frac{I_1^{(N)}(r_{b1},e^{\imath\psi}r_{a2})}{r_{b1}-e^{\imath\psi}r_{a2}}\nonumber\\
 \fl&=&\frac{(2\pi)^2\imath}{(N-1)!}\frac{r_{b1}^N\Theta(r_{b1})}{r_{b1}-e^{\imath\psi}r_{a2}}\left(-e^{-\imath\psi}\frac{\partial}{\partial r_{a2}}\right)^{N-1}e^{-\imath\psi}\delta(r_{a2})\,,\label{a7.4}\\
 \fl A_{b2}(e^{\imath\psi}r_{a2})&=&\exp\left(-\varepsilon e^{\imath\psi}r_{a2}\right)\int\limits_{\mathbb{R}}\exp\left(\imath s_{2}r_{a2}\right)\left(e^{-\imath\psi}s_{2}+\imath\varepsilon\right)^{b-1}e^{-\imath\psi}ds_2\prod\limits_{j=1}^k\chi(r_{j1}-e^{\imath\psi}r_{a2})\nonumber\\
 \fl &=&2\pi\left(-\imath e^{-\imath\psi}\frac{\partial}{\partial r_{a2}}\right)^{b-1}e^{-\imath\psi}\delta(r_{a2})\prod\limits_{j=1}^k\chi(r_{j1}-e^{\imath\psi}r_{a2})\,,\label{a7.5}\\
 \fl A_3(r_{b1},\alpha E_a^{(0)})&=&\exp\left(\varepsilon r_{b1}\right)\int\limits_{\mathbb{R}}\frac{\exp\left(-\imath s_{1}r_{b1}\right)}{s_{1}+\imath\varepsilon+\alpha E_{a}^{(0)}}\left(\frac{-\alpha E_{a}^{(0)}}{s_{1}+\imath\varepsilon}\right)^Nds_1\prod\limits_{j=1}^k\chi(r_{b1}-e^{\imath\psi}r_{j2})\nonumber\\
 \fl &=&2\pi\imath\Theta(r_{b1})\sum\limits_{n=N}^\infty\frac{(\imath \alpha E_{a}^{(0)} r_{b1})^n}{n!}\prod\limits_{j=1}^k\chi(r_{b1}-e^{\imath\psi}r_{j2})\,.\label{a7.6}
\end{eqnarray}
Surprisingly this part of our result agrees with the one in Ref.~\cite{KieGuh09a} (apart from a forgotten $2\pi\imath$ in the upper left block of Eq.~(6.9) in Ref.~\cite{KieGuh09a} and the characteristic functions $\chi(r_{b1}-e^{\imath\psi}r_{a2})$) although we omitted all Efetov-Wegner terms in this work.

The second step is to diagonalize the supermatrix $\rho$ in Eq.~\eref{6.2.1},
\begin{eqnarray}
 \fl&&Z_{k}^{(N)}(\kappa,\alpha H_0)\label{a7.7}\\
 \fl&=&(-\imath)^{k(k+1)/2}\int\limits_{\mathbb{R}^{2k}}\det\left[\displaystyle \frac{-2\pi e^{-\imath\psi}\delta(r_{b1})\delta(r_{a2})}{\kappa_{b1}-\kappa_{a2}}+\frac{\exp\left(-\imath\kappa_{b1}r_{b1}+\imath e^{\imath\psi}\kappa_{a2}r_{a2}\right)}{r_{b1}-e^{\imath\psi}r_{a2}}\chi(\kappa_{b1}-\kappa_{a2})\right]_{1\leq a,b\leq k}\nonumber\\
 \fl&\times&\hspace*{-0.15cm}\det\hspace*{-0.1cm}\left[\begin{array}{c|c}\displaystyle-\frac{I_1^{(N)}(r_{b1},e^{\imath\psi}r_{a2})}{2\pi(r_{b1}-e^{\imath\psi}r_{a2})} & \displaystyle\left(e^{-\imath\psi}\frac{\partial}{\partial r_{a2}}\right)^{b-1}e^{-\imath\psi}\delta(r_{a2})\prod\limits_{j=1}^k\chi(r_{j1}-e^{\imath\psi}r_{a2}) \\ \hline \displaystyle\Theta(r_{b1})\sum\limits_{n=N}^\infty\frac{(\imath \alpha E_{a}^{(0)} r_{b1})^n}{n!}\prod\limits_{j=1}^k\chi(r_{b1}-e^{\imath\psi}r_{j2}) & \displaystyle(-\imath\alpha E_{a}^{(0)})^{b-1} \end{array}\right]\nonumber\\
 \fl&\times&\frac{\Phi(r)d[r]}{\Delta_N(\alpha E_0)\sqrt{\Ber_{k/k}^{(2)}(\kappa)}\sqrt{\Ber^{(2)}_{k/k}(r)}}.\nonumber
\end{eqnarray}
The range of the indices in the second determinant is the same as in Eq.~\eref{a7.3}. Expanding the first determinant we have
\begin{eqnarray}
 \fl&&Z_{k}^{(N)}(\kappa,\alpha H_0)\label{a7.8}\\
 \fl&=&\frac{(-\imath)^{k(k+1)/2}}{\Delta_N(\alpha E_0)\sqrt{\Ber_{k/k}^{(2)}(\kappa)}}\sum\limits_{l=0}^{k}\sum\limits_{\omega_1,\omega_2\in\mathfrak{S}(k)}\frac{\sign\omega_1\omega_2}{\left[l!(k-l)!\right]^2}\prod\limits_{a,b=l+1}^k\chi(\kappa_{\omega_1(b)1}-\kappa_{\omega_2(a)2})\nonumber\\
 \fl&\times&\int\limits_{\mathbb{R}^{2k}}\det\left[\displaystyle \frac{\exp\left(-\imath\kappa_{\omega_1(b)1}r_{\omega_1(b)1}+\imath e^{\imath\psi} \kappa_{\omega_2(a)2}r_{\omega_2(a)2}\right)}{r_{\omega_1(b)1}-e^{\imath\psi}r_{\omega_2(a)2}}\right]_{l+1\leq a, b\leq k}\nonumber\\
 \fl&\times&\hspace*{-0.15cm}\det\hspace*{-0.1cm}\left[\begin{array}{c|c}\displaystyle-\frac{I_1^{(N)}(r_{b1},e^{\imath\psi}r_{a2})}{2\pi(r_{b1}-e^{\imath\psi}r_{a2})} & \displaystyle\left( e^{-\imath\psi}\frac{\partial}{\partial r_{a2}}\right)^{b-1}e^{-\imath\psi}\delta(r_{a2})\prod\limits_{j=1}^k\chi(r_{j1}-e^{\imath\psi}r_{a2}) \\ \hline \displaystyle\Theta(r_{b1})\sum\limits_{n=N}^\infty\frac{(\imath \alpha E_{a}^{(0)} r_{b1})^n}{n!}\prod\limits_{j=1}^k\chi(r_{b1}-e^{\imath\psi}r_{j2}) & \displaystyle(-\imath\alpha E_{a}^{(0)})^{b-1} \end{array}\right]\nonumber\\
 \fl&\times&\det\left[\frac{-2\pi e^{-\imath\psi}\delta(r_{\omega_1(b)1})\delta(r_{\omega_2(a)2})}{\kappa_{\omega_1(b)1}-\kappa_{\omega_2(a)2}}\right]_{1\leq a,b\leq l}\frac{\Phi(r)d[r]}{\sqrt{\Ber^{(2)}_{k/k}(r)}}\,.\nonumber
\end{eqnarray}
When integrating and summing up we use the normalization~\eref{a6.2} of the supersymmetric Ingham-Siegel integral and arrive at the result~\eref{6.2.2}.

\section*{References}


\end{document}